\RequirePackage{fix-cm} 
\documentclass[a4paper, twoside, reqno, 12pt, dvips]{amsart}

\usepackage{fixltx2e}     

\usepackage[light, firsttwo, outline]{draftcopy}

\usepackage{amssymb}
\usepackage{amsmath}

\usepackage{a4wide}

\usepackage{indentfirst}
\usepackage{graphicx}
\usepackage{psfrag}

\usepackage[usenames,dvipsnames]{pstricks}
\usepackage{epsfig}
\usepackage{pst-grad} 
\usepackage{pst-plot} 

\usepackage{subfigure}

\usepackage{longtable}

\usepackage[english]{babel}
\usepackage[latin1]{inputenc}

\usepackage{color}
\definecolor{oneblue}{rgb}{0,0.0,0.75}
\usepackage[colorlinks,
            urlcolor=oneblue,
            linkcolor=oneblue,
            citecolor=oneblue,
            bookmarksopen=false,
            pagebackref]{hyperref}

\numberwithin{equation}{section}

\newtheorem{remark}{Remark}

\newcommand{\x}{\vec{x}}
\newcommand{\n}{\hat{n}}
\renewcommand{\k}{\vec{k}}
\renewcommand{\u}{\vec{u}}
\newcommand{\grad}{\nabla}
\newcommand{\phis}{\varphi}
\newcommand{\R}{\mathbb{R}}
\newcommand{\D}{\mathcal{D}}
\newcommand{\Hv}{\mathcal{H}}

\renewcommand{\O}{\mathcal{O}}

\newcommand{\F}{\mathcal{F}}
\newcommand{\dz}{\partial_z}
\newcommand{\dt}{\partial_t}

\renewcommand{\div}{\nabla\cdot}

\newcommand{\half}{{\textstyle{1\over2}}}
\newcommand{\sech}{\mathop{\mathrm{sech}}}
\newcommand{\pd}[2]{\frac{\partial#1}{\partial#2}}

\begin{document}

\title[Contribution of horizontal displacements into tsunami generation]{On the contribution of the horizontal sea-bed displacements into the tsunami generation process}

\author[D. Dutykh]{Denys Dutykh$^*$}
\address{LAMA, UMR 5127 CNRS, Universit\'e de Savoie, Campus Scientifique,
73376 Le Bourget-du-Lac Cedex, France}
\email{Denys.Dutykh@univ-savoie.fr}
\urladdr{http://www.lama.univ-savoie.fr/~dutykh/}
\thanks{$^*$ Corresponding author. Tel: +33 4 79 75 86 52, Fax: +33 4 79 75 81 42. E-mail: \texttt{Denys.Dutykh@univ-savoie.fr}}

\author[D. Mitsotakis]{Dimitrios Mitsotakis}
\address{IMA, University of Minnesota, 114 Lind Hall,
207 Church Street SE, Minneapolis MN 55455, USA}
\email{dmitsot@gmail.com}
\urladdr{http://sites.google.com/site/dmitsot/}

\author[L. B. Chubarov]{Leonid B. Chubarov}
\address{Institute of Computational Technologies, Siberian Branch of the Russian Academy of Sciences, 6 Acad. Lavrentjev Avenue, 630090 Novosibirsk, Russia}
\email{chubarov@ict.nsc.ru}
\urladdr{http://www.ict.nsc.ru/}

\author[Yu. I. Shokin]{Yuriy I. Shokin}
\address{Institute of Computational Technologies, Siberian Branch of the Russian Academy of Sciences, 6 Acad. Lavrentjev Avenue, 630090 Novosibirsk, Russia}
\email{shokin@ict.nsc.ru}
\urladdr{http://www.ict.nsc.ru/}

\begin{abstract}
The main reason for the generation of tsunamis is the deformation of the bottom of the ocean caused by an underwater earthquake. Usually, only the vertical bottom motion is taken into account while the horizontal co-seismic displacements are neglected in the absence of landslides. In the present study we propose a methodology based on the well-known Okada solution to reconstruct in more details all components of the bottom coseismic displacements. Then, the sea-bed motion is coupled with a three-dimensional weakly nonlinear water wave solver which allows us to simulate a tsunami wave generation. We pay special attention to the evolution of kinetic and potential energies of the resulting wave while the contribution of the horizontal displacements into wave energy balance is also quantified. Such contribution of horizontal displacements to the tsunami generation has not been discussed before, and it is different from the existing approaches. The methods proposed in this study are illustrated on the July 17, 2006 Java tsunami and some more recent events.
\end{abstract}


\maketitle

\tableofcontents

\section{Introduction}

During the years 2004 to 2006 several interesting tsunami events took place in the Indian Ocean. In December 26, 2004, the Great Sumatra-Andaman earthquake ($M_w = 9.1$, cf. C.~Ammon \emph{et al.} (2005), \cite{Ammon2005}) generated the devastating Indian Ocean tsunami refered to in the literature as the Tsunami Boxing Day 2004 (cf. C.~Synolakis \& E.~Bernard (2006), \cite{Syno2006}). The local tsunami run-ups from this event exceeded the height of 34 m at Lhoknga in the western Aceh Province. This and other observations led many researchers to ask whether this tsunami was unusualy large for this specific earthquake size. Later it was shown by E.L.~Geist \emph{et al.} (2006) \cite{Geist2006a} that the Great Sumatra-Andaman earthquake is very similar in terms of local tsunami magnitude to past events of the same size. For example, the 1964 Great Alaska earthquake ($M_w = 9.2$, cf. H.~Kanamori (1970) \cite{Kanamori1970}) demonstrates a similar scaling.

On March 28, 2005 another earthquake occured approximately 110 $km$ to the SE from the Great Sumatra-Andaman earthquake's epicenter. The magnitude of this earthquake was estimated to be $M_w = 8.6 \sim 8.7$ (cf. \cite{Lay, Bilham2005, Walker2005}). This event triggered a massive evacuation in the surrounding Indian Ocean countries. However, the March 28, 2005 Northern Sumatra earthquake failed to generate a significant tsunami event. The survey teams reported maximum tsunami run-up of 4 $m$. \cite{Geist2006a}. This event can be compared with the 1957 Aleutian earthquake ($M_w = 8.6$, cf. J.M.~Johnson and K.~Satake (1993) \cite{Johnson1993}) which produced a maximum tsunami run-up of 15 $m$ (see J.F.~Lander (1996) \cite{Lander1996}). The deficiency of the March 2005 tsunami is related to the slip concentration in the down-dip part of the rupture zone and to the fact that a substantial part of the vertical displacement occurred in shallow waters or on the substance of the ground \cite{Geist2006a}.

On the other hand, the smaller July 16, 2006 Java earthquake ($M_w = 7.8$, cf. C.J.~Ammon \emph{et al.} (2006), \cite{Ammon2006}) generated an unexpectedly destructive tsunami wave which affected over 300 km of south Java coastline and killed more than 600 people, \cite{Fritz2007}. Field measurements of run-up distributions range uniformly from 5 to 7 $m$ in most inundated areas. However, unexpectedly high run-up values  were reported at Nusakambangan Island exceeding the value of 20 $m$. This tsunami focusing effect could be seemingly ascribed to local site effects and/or to a local submarine landslide/slump. July 16, 2006 Java tsunami can be compared to a similar event occured on June 2, 1994 at the East Java ($M_w = 7.6$) (see Y.~Tsuji \emph{et al.} (1995), \cite{Tsuji1995}). This 1994 Java tsunami produced more than 200 casualties with local run-up at Rajekwesi slightly exceeding 13 $m$.

All these examples of recent and historical tsunami events show that there is a big variety of the local tsunami heights and run-up values with respect to the earthquake magnitude $M_w$. It is obvious that the seismic moment $M_0$ of underwater shallow earthquakes is adequate to estimate far-field tsunami amplitudes (see also \cite{Okal2003b}). However, tsunami wave energy reflects better the local tsunami severity, while the specific run-up distribution depends on bathymetric propagation paths and other site-specific effects. One of the central challenges in the tsunami science is to rapidly assess a local tsunami severity from very first rough earthquake estimations. In the current state of our knowledge false alarms for local tsunamis appear to be unavoidable. The tsunami generation modeling attempts to improve our understanding of tsunami behaviour in the vicinity of the genesis region.

Tsunami generation modeling was initiated in the early sixties by the prominent work of K.~Kajiura, \cite{kajiura}, who proposed the use of the static vertical sea bed displacement for the initial condition of the free surface elevation. Classically, the celebrated Okada, \cite{Okada85, okada92}, and sometimes Mansinha \& Smylie\footnote{In fact, Mansinha \& Smylie solution is a particular case of the more general Okada solution.} \cite{Mansinha1967, Mansinha1971} or Gusiakov,\cite{Gusiakov1978a}, solutions are used to compute the coseismic sea bed displacements. This approach is still widely used by the tsunami wave modeling community. However, some progress has been recently made in this direction,\cite{Ohmachi2001, Dutykh2006, Dutykh2007a, Dutykh2007b, Rabinovich2008, Saito2009, Dutykh2009a}.

There is a consensus on the importance of the horizontal motion in landslide generated tsunamis, \cite{Harbitz1992, Ward, Bardet2003, Okal2003}. However, for tsunami waves caused by underwater earthquakes, the horizontal displacements are often disregarded by the tsunami wave modeling community. We can quote here a few publications devoted to tsunami waves such as one by D.~Stevenson (2005), \cite{Stevenson2005}:
\begin{quote}
	``Although horizontal displacements are often larger, they are unimportant for tsunami generation except to the extent that the sloping ocean floor also forces a vertical displacement of the water column.''
\end{quote}
Or another one by G.A.~Ichinose {\em et al.} (2000) \cite{Ichinose2000}:
\begin{quote}
  ``The initial lake level values were specified by assuming that the lake surface instantaneously conformed to the vertical displacement of the lake bottom while horizontal velocities were set to zero. The effect of horizontal deformation on the initial condition is neglected here and left for future work.''
\end{quote}
The authors of this article also tended to neglect the horizontal displacements field in previous tsunami generation studies, \cite{Dutykh2007a, Mitsotakis2007}. Perhaps, this situation can be ascribed to the work of E.~Berg (1970), \cite{Berg1970}, who showed in the case of the 1964 Alaska's earthquake ($M_w = 9.2$) that the input into the potential energy from the horizontal motion is less than 1.5\% of that from the vertical movement.

The attitude to horizontal displacements changed after the prominent work by Y.~Tanioka and K.~Satake (1996), \cite{Tanioka1996}. According to their computations for the 1994 Java earthquake, the inclusion of horizontal displacements contribution results in 43\% increase in maximum initial vertical ground displacement and an increase in the wave amplitudes at the shoreline by 30\%. These striking results incited researchers to reconsider the role of the horizontal sea bed motion. Hereafter, various tsunami generation techniques which involve only the vertical displacement field are referred to as the incomplete scenario. On the contrary, the complete tsunami generation takes also into account the horizontal displacements field. The term \emph{complete} in our study refers to a purely kinematic interaction between the moving bottom and the ocean layer in contrast to the horizontal impulse-momentum transfer mechanism \cite{Vennard1982} applied for the first time to tsunami wave generation in \cite{Song2008}.

The approach proposed by Tanioka \& Satake (1996) is based on a simple physical consideration: horizontal displacements of a sloping bottom will produce some amount of the vertical motion depending on the slope magnitude. Assuming that the slope is small, this idea can be expressed mathematically by applying once the first order Taylor expansion:
\begin{equation*}
  u_h := u_1 h_x + u_2 h_y,
\end{equation*}
where $u_{1,2}$ are the horizontal components of the displacements field and $h(x,y)$ is the bathymetry function. The indices $x$ and $y$ denote the spatial partial derivatives. Tanioka \& Satake (1996) referred to the quantity $u_h$ as the \emph{vertical displacement of water due to the horizontal movement of the slope}. Then, the quantity $u_h$ is simply added as a correction to the vertical displacements component $u_3(x,y)$. To compute the co-seismic displacements $u_{1,2,3}$ they used the celebrated Okada solution \cite{stek2, Okada85, okada92} applied to a single rectangular fault segment. In the present study we apply the same solution to multiple sub-faults to achieve higher resolution (see also \cite{Dutykh2010a}). More recently, an original approach based on the impulse-momentum principle, \cite{Vennard1982}, has been proposed in \cite{Song2008}. Their genesis theory is based on the idea of the momentum exchange between the slipped continental slope and the water column due to the earthquake. The horizontal bottom motion contributes directly to the ocean kinetic energy as any moving object in water transfers momentum to the fluid. This idea is similar to the classical added mass concept. According to the authors of \cite{Song2008}, the continental slope is not an exception from this principle. This implies some special treatment of the water parcels in the vicinity of moving boundaries. For more details we refer to the original publication \cite{Song2008}. The horizontal impulse approach leads to predictions which are somehow different from the present and several previous studies. The methodology described in this article is closer to the ideas of Tanioka \& Satake (1996) \cite{Tanioka1996}. Namely, we reconstruct in more details the kinematics of bottom displacements in vertical and horizontal directions, including their evolution in time. We underline that most of tsunami generation studies, including \cite{Tanioka1996}, assume the bottom deformation process to be instantaneous, that in most of the cases provides a fairly good approximation. Then, the bottom motion generates some perturbations on the free surface of the water layer. This perturbation form long waves, propagating across the oceans under the gravity force and commonly referred to as \emph{tsunami waves} due to their destructive potential fully realized only in coastal regions, \cite{kajiura, Velichko2002, Dias2006, Syno2006}.

In the current study we shed some light into the energy transfer process from an underwater earthquake to the implied tsunami wave (in the spirit of the study by D.~Dutykh \& F.~Dias (2009), \cite{Dutykh2009b}), while taking into account and quantifying the horizontal displacements contribution into tsunami energy balance. We focus only on the generation stage since the propagation phase and run-up techniques are well understood nowadays at least in the context of Nonlinear Shallow Water equations, \cite{Imamura1996, DeKaKa, Kim2007, Dutykh2009a, Dutykh2010}.

The present study is organized as follows. In Section \ref{sec:models} we present mathematical models used in this study. Specifically, in Sections \ref{sec:displs} and \ref{sec:wnsolver} a description of the bottom motion and the fluid layer solution is presented respectively. Some rationale on tsunami wave energy computations is discussed in Section \ref{sec:energy}. Numerical results are presented in Section \ref{sec:numres}. Finally, some important conclusions of this study are outlined in Section \ref{sec:concl}. In the Appendix of this paper we present the results concerning the tsumami generation of two recent events.

\section{Mathematical models}\label{sec:models}

Tsunami waves are caused by a huge and rapid motion of the seafloor due to an underwater earthquake over broad areas in comparison to water depth. There are some other mechanisms of tsunami genesis such as underwater landslides, for example. However, in this study we focus on the purely seismic mechanism which occurs most frequently in nature.

Hydrodynamics and seismology are only weakly coupled in the tsunami generation problem. Namely, the released seismic energy is partially transmitted to the ocean through the sea bed deformation while the ocean has obviously no influence onto the rupturing process. Consequently, our problem is reduced to two relatively distinct questions:
\begin{enumerate}
  \item Reconstruct the sea bed deformation $h=h(\x,t)$ caused by the seismic event under consideration
  \item Compute the resulting free surface motion
\end{enumerate}
The answers to these questions are analyzed in Sections \ref{sec:displs} and \ref{sec:wnsolver} respectively.

\subsection{Dynamic bottom displacements reconstruction}\label{sec:displs}

Traditionally, the free surface initial condition for various tsunami propagation codes (see \cite{Titov1997, Imamura1996, Imamura2006}) is assumed to be identical to the static vertical deformation of the ocean bottom, \cite{Kajiura1970}. This assumption is classically justified by the three following reasons:
\begin{enumerate}
  \item Tsunamis are long waves. In this regime the vertical acceleration is neglected with respect to the gravity force
  \item The sea bed deformation is assumed to be instantaneous. It is based on the comparison of gravity wave speed ($200$ m/s for water depth of $4$ km) and the seismic wave celerity ($\approx 3600$ m/s)
  \item The effect of horizontal bottom motion is negligible for tsunami generation since the bathymetry has in general mild slope ($\approx 10\%$), \cite{Berg1970}
\end{enumerate}
It is worth to note that nonhydrostatic effects as well as finite time source duration have been modeled in several recent studies \cite{Todo, Dutykh2006, Fructus2007, Kervella2007, Dutykh2007b, Fuhrman2009}. Some attempts have also been made to take into account the horizontal displacements contribution, \cite{Tanioka1996, Geist2006a, Song2008}.

In the present study we relax all three classical assumptions. The bottom deformation is reconstructed using the finite fault solution as it was suggested in our previous study \cite{Dutykh2010a}. However, we extend the previous construction to take into account the horizontal displacements contribution as well. The finite fault solution is based on the multi-fault representation of the rupture, \cite{Bassin2000, Ji2002}. The rupture complexity is reconstructed using the inversion of seismic data. Fault's surface is parametrized by multiple segments with variable local slip, rake angle, rise time and rupture velocity. The inversion is performed in an appropriate wavelet transform space. The objective function is a weighted sum of $L_1$, $L_2$ norms and some correlative functions. With this approach seismologists are able to recover rupture slip details, \cite{Bassin2000, Ji2002}. This available seismic information is exploited hereafter to compute the sea bed displacements produced by an underwater earthquake with better geophysical resolution. Several other multiple segments sources have been used in \cite{Geist2002, Wang2006, McCloskey2008}.

\begin{remark}
In a few studies an attempt has been made to reconstruct the seismic source from tsunami tide gauge records, \cite{Piatanesi2001, Pires2003, Yagi2004, Fujii2006, Ichinose2007, Rhie2007}. This approach seems to be very promising and in future a joint combination of seismic and hydrodynamic inversions should be used for the successful reconstruction of appropriate initial conditions.
\end{remark}

Let us describe the way of how the sea bed displacements are reconstructed. In this reconstruction procedure we follow the great lines of our previous study \cite{Dutykh2010a}, while adding new ingredients concerning the horizontal displacements field reconstruction. We illustrate the proposed approach in the case of the July 17, 2006 Java tsunami for which the finite fault solution is available, cf. \cite{Ji2006, Ozgun2006}. It was a relatively slow earthquake and thus, atypical. However, we assume that the slow rupturing process was well resolved by the finite fault inversion algorithm.

\begin{figure}
  \centering
  \includegraphics[scale=0.6]{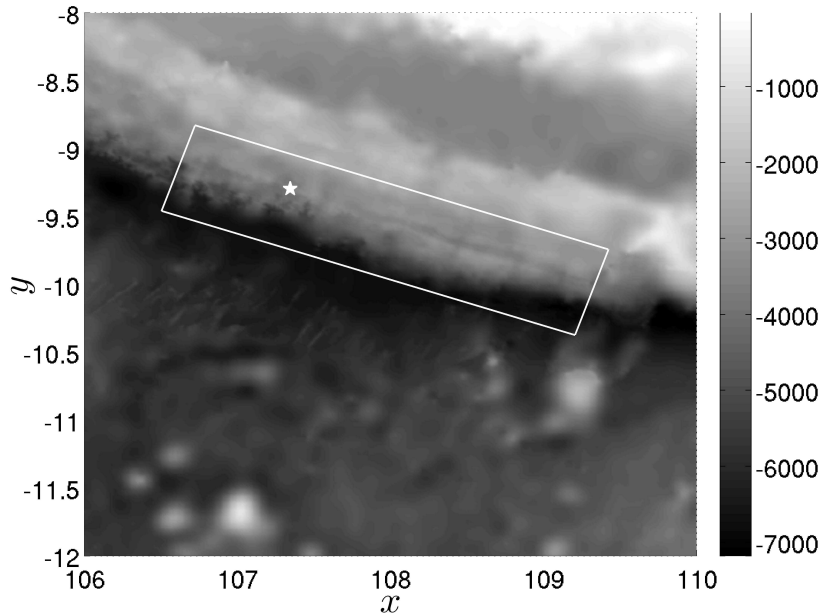}
  \caption{Surface projection of the fault's plane and the ETOPO1 bathymetric map of the region under consideration. The symbol $\star$ indicates the hypocenter's location at $(107.345^\circ, -9.295^\circ)$. The local Cartesian coordinate system is centered at the point $(108^\circ, -10^\circ)$. This region is located between $(106^\circ, -8^\circ)$ and $(110^\circ, -12^\circ)$. The colorbar indicates the water depth in meters below the still water level ($z=0$). In the region under consideration the depth varies from 20 to 7100 meters.}
  \label{fig:bathy1}
\end{figure}

The fault is considered to be the rectangle with vertices located at $(109.20508^\circ$ (Lon), $-10.37387^\circ$ (Lat), $6.24795$ $km$ (Depth)$)$, $(106.50434^\circ$, $-9.45925^\circ$, $6.24795\;km)$, $(106.72382^\circ$, $-8.82807^\circ$, $19.79951\;km)$, $(109.42455^\circ$, $-9.74269^\circ$, $19.79951$ $km)$ (see Figure \ref{fig:bathy1}). The fault's plane is conventionally divided into $N_x = 21$ subfaults along strike and $N_y = 7$ subfaults down the dip angle, leading to the total number of $N_x\times N_y = 147$ equal segments. Parameters such as the subfault location $(x_c, y_c)$, the depth $d_i$, the slip $u_i$ and the rake angle $\phi_i$ for each segment can be found in \cite{Ji2006} (see also Appendix II, \cite{Dutykh2010a}). The elastic constants and parameters such as dip and slip angles, which are common to all subfaults, are given in Table \ref{tab:crust}. We underline that the slip angle is measured conventionally in the counter-clockwise direction from the North. The relations between the elastic wave celerities $c_p$, $c_s$ and Lam\'e coefficients $\lambda$, $\mu$ used in Okada's solution are classical and can also be found in Appendix III, \cite{Dutykh2010a}.

\begin{table}
\begin{center}
\begin{tabular}{c||c}
  \hline\hline
  $P$-wave celerity $c_p$, $m/s$ & 6000 \\
  \hline
  $S$-wave celerity $c_s$, $m/s$ & 3400 \\
  \hline
  Crust density $\rho$, $kg/m^3$ & 2700 \\
  \hline
  Dip angle, $\delta$ & $10.35^\circ$ \\
  \hline
  Slip angle (CW from N) & $288.94^\circ$ \\
  \hline\hline
\end{tabular}
\caption{Geophysical parameters used to model elastic properties of the subduction zone in the region of Java.}
\label{tab:crust}
\end{center}
\end{table}

One of the main ingredients in our construction is the so-called Okada solution, \cite{Okada85, okada92}, which is used in the case of an active fault of small or intermediate size. The success of this solution may be ascribed to the closed-form analytical expressions which can be effectively used for various modeling purposes involving co-seismic deformation.

\begin{remark}
The celebrated Okada solution, \cite{Okada85, okada92}, is based on two main ingredients --- the dislocation theory of Volterra \cite{volt, stek2} and Mindlin's fundamental solution for an elastic half-space, \cite{mindl1}. Particular cases of this solution were known before Okada's work, for example the well-known Mansinha and Smylie's solution, \cite{Mansinha1967, Mansinha1971}. Usually all these particular cases differ by the choice of the dislocation and the Burger's vector orientation, \cite{press}. We recall the basic assumptions behind this solution:
\begin{itemize}
  \item Fault is immersed into the linear homogeneous and isotropic half-space
  \item Fault is a Volterra's type dislocation
  \item Dislocation has a rectangular shape
\end{itemize}
For more information on Okada's solution we refer to \cite{Dutykh2006, Dias2006, Dutykh2007a} and the references therein.
\end{remark}

The trace of the Okada's solution at the sea bottom (substituting $z=0$ in the geophysical coordinate system) for each subfault will be denoted $\O_i^{(j)}(\x; \delta, \lambda, \mu, \ldots)$, where $\delta$ is the dip angle, $\lambda$, $\mu$ are the Lam\'e coefficients and dots denote the dependence of the function $\O_i^{(j)}(\x)$ on other 8 parameters, cf. \cite{Dutykh2006}. The index $i$ takes values from $1$ to $N_x\times N_y$ and denotes the corresponding subfault segment, while the superscript $j$ is equal to $1$ or $2$ for the horizontal displacements and to $3$ for the vertical component. Hereafter we will adopt the short-hand notation $\O_i^{(j)}(\x)$ for the $j^{\mathrm{th}}$ displacement component of the Okada's solution for the $i^{\mathrm{th}}$ segment having in mind its dependence on various parameters.

\begin{figure}
  \centering
    \includegraphics[width=0.79\textwidth]{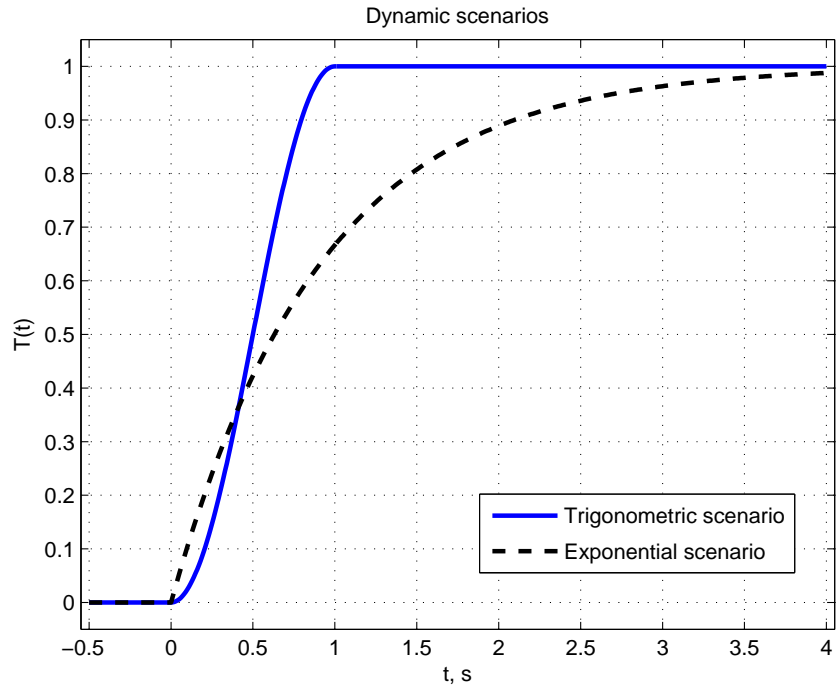}
  \caption{Trigonometric and exponential dynamic scenarios for $t_r = 1$ s (see J.~Hammack (1973), \cite{Hammack}).}
  \label{fig:scenario}
\end{figure}

Taking into account the dynamic characteristics of the rupturing process, we make some further assumptions on the time dependence of the displacement fields. The finite fault solution provides us with two additional parameters concerning the rupture kinematics --- the rupture velocity $v_r$ and the rise time $t_r$ which are equal to $1.1$ km/s and $8$ s for July 17, 2006 Java event respectively. The epicenter is located at the point $\x_e =$ $(107.345^\circ, -9.295^\circ)$, cf. \cite{Ji2006}. Given the origin $\x_e$, the rupture velocity $v_r$ and $i^{\mathrm{th}}$ subfault location $\x_i$, we define the {\em subfault activation times} $t_i$ needed for the rupture to achieve the corresponding segment $i$ by the formulas:
\begin{equation*}
  t_i = \frac{||\x_e - \x_i ||}{v_r}, \quad i=1,\ldots, N_x\times N_y.
\end{equation*}
Consequently, for the sake of simplicity in our study we assume the rupture propagation along the fault to be isotropic and homogeneous. However, some more sophisticated approaches including possible heterogeneities exist (see e.g. \cite{Das1996}). We will also follow the pioneering idea of J.~Hammack, \cite{Hammack1972, Hammack}, developed later in \cite{Todo, todo2, Dutykh2006, ddk, Kervella2007}, where the maximum bottom deformation is achieved during some finite time (the so-called {\em rise time}) according to an appropriately chosen dynamic scenario. Various scenarios used in practice (instantaneous, linear, trigonometric, exponential, etc) can be found in \cite{Hammack1972, Kanamori1972, Hammack, ddk, Dutykh2006}. In this study we adopt the trigonometric scenario which is given by the following formula:
\begin{equation}\label{eq:expscenario}
  T(t) = \Hv(t-t_r) + \frac12\Hv(t)\Hv(t_r-t)\bigl(1 - \cos(\pi t/t_r)\bigr),
\end{equation}
where $\Hv(t)$ is the Heaviside step function. This scenario has the advantage to have also the first derivative continuous at the activation time $t = 0$. However, for comparative purposes sometimes we will use also the so-called exponential scenario \cite{Kanamori1972}:
\begin{equation*}
  T_e(t) = \Hv(t)\bigl(1 - e^{-\alpha t}\bigr), \qquad \alpha := \frac{\log(3)}{t_r}.
\end{equation*}
For illustrative purposes both dynamic scenarios are represented in Figure \ref{fig:scenario}.

We sum up together all the ingredients proposed above to reconstruct dynamic displacements field $\u = (u_1, u_2, u_3)$ at the sea bottom:
\begin{equation*}
  u_j(\x,t) = \sum_{i=1}^{N_x\times N_y}T(t-t_i) \O_i^{(j)}(\x).
\end{equation*}

\begin{remark}
We would like to underline here the asymptotic behaviour of the sea bed displacements. By definition of the trigonometric scenario \eqref{eq:expscenario} we have $\lim\limits_{t\to +\infty}T(t) = 1$. Consequently, the sea bed deformation will attain fast its state which consists of the linear superposition of subfaults contributions:
\begin{equation*}
  u_j(\x,t) = \sum_{i=1}^{N_x\times N_y} \O_i^{(j)}(\x).
\end{equation*}
\end{remark}

Finally, we can predict the sea bed motion by taking into account horizontal and vertical displacements:
\begin{equation}\label{eq:bathym}
  h(\x, t) = h_0 (\x - \u_{1,2}(\x, t)) - u_3 (\x, t),
\end{equation}
where $h_0(\x)$ is a function which interpolates\footnote{In our numerical simulations presented below we use the MATLAB \texttt{TriScatteredInterp} class to interpolate the static bathymetry values given by ETOPO1 database.} the static bathymetry profile given e.g. by the ETOPO1 database (see Figures \ref{fig:bathy1} and \ref{fig:bathy2}).

\begin{remark}
In some studies where horizontal displacements were taken into account (cf. \cite{Tanioka1996, Baba2006, Song2008, Beisel2009}), the first order Taylor expansion was permanently applied to the bathymetry representation formula \eqref{eq:bathym} to give:
\begin{equation}\label{eq:taylor}
  h(\x, t) \approx h_0(\x) - \u_{1,2}(\x,t)\cdot\grad_{\x}h_0(\x) - u_3(\x, t)
\end{equation}
We prefer not to follow this approximation and to use the exact formula \eqref{eq:bathym} since it is valid for all slopes (see Figure \ref{fig:bathy2} for Java bathymetry). Another difficulty lies in the estimation of the bathymetry gradient $\grad_{\x}h_0(x)$ required by Taylor's formula \eqref{eq:taylor}. The application of any finite differences scheme to the measured $h_0(\x)$ leads to an ill-posed problem. Consequently, one needs to apply extensive smoothing procedures to the raw data $h_0(\x)$ which induces an additional loss in accuracy.

In the present study we do not completely avoid the computation of the static bathymetry gradient $\grad_{\x}h_0(\x)$ since the kinematic bottom boundary condition \eqref{eq:bottomkin} involves the time derivative of the bathymetry function:
\begin{equation*}
  \dt h = -\grad_{\x}h_0(\x)\cdot\dt\u_{1,2}(\x,t) - \dt u_3(\x, t).
\end{equation*}
However, the last formula is exact and it is obtained by a straightforward application of the chain differentiation rule.

Another possibility could be to consider static horizontal displacements $\u_{1,2}(\x)$ thus keeping dynamics only in the vertical component $u_3(\x, t)$. However, we do not choose this option in this work.
\end{remark}

In the next section we will present our approach in coupling this dynamic deformation with the hydrodynamic problem to predict waves induced on the free surface of the ocean.

\subsection{The water wave problem with moving bottom}\label{sec:wnsolver}

\begin{figure}
  \centering
  \includegraphics[width=0.9\textwidth]{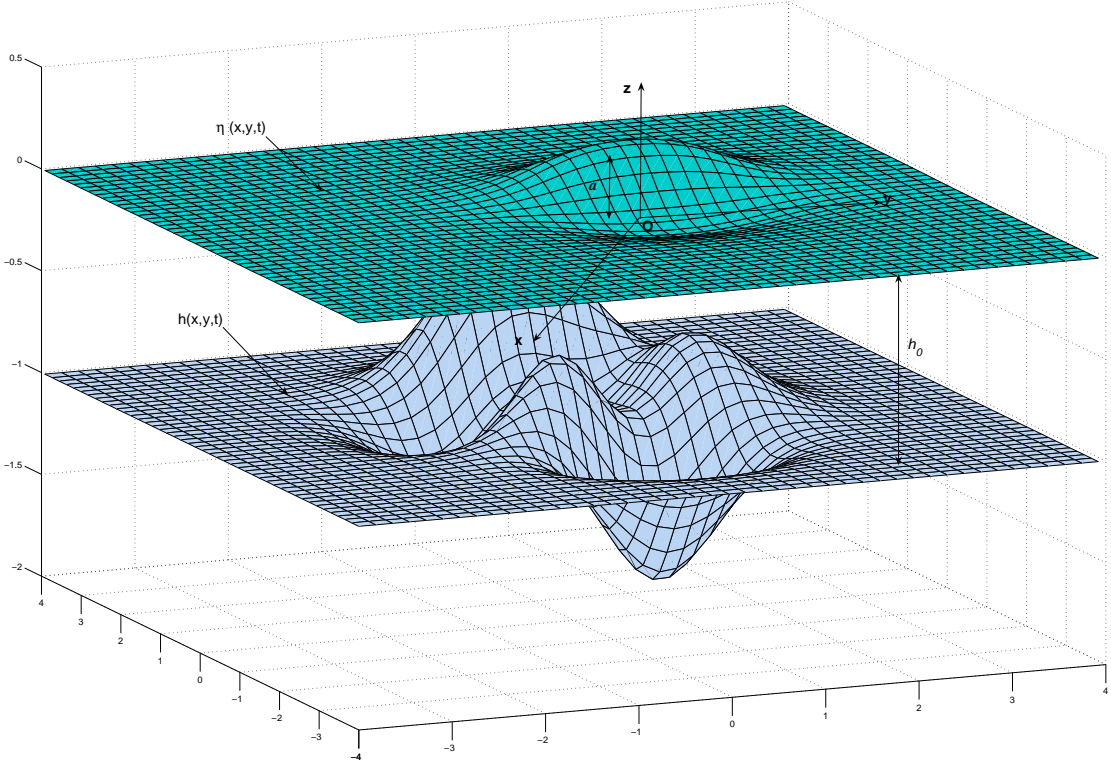}
  \caption{Sketch of the physical domain.}
  \label{fig:sketch}
\end{figure}

We consider the incompressible flow of an ideal fluid with constant density $\rho$ in the domain $\Omega\subseteq\R^2$. The horizontal independent variables will be denoted by $\x = (x,y)$ and the vertical one by $z$. The origin of the Cartesian coordinate system is traditionally chosen such that the surface $z=0$ corresponds to the still water level. The fluid domain is bounded below by the bottom $z = -h(\x,t)$ and above by the free surface $z = \eta (\x,t)$. Usually we assume that the total depth $H(\x, t) := h(\x,t) + \eta (\x,t)$ remains positive $H (\x,t) \geq h_0 > 0$ under the system dynamics $\forall t \in [0, T]$. The sketch of the physical domain is shown in Figure \ref{fig:sketch}.

\begin{remark}
Classically in water wave modeling, we make the assumption that the free surface is a graph $z = \eta (\x,t)$ of a single-valued function. It means in practice that we exclude some interesting phenomena, (e.g. wave breaking phenomena) which are out of the scope of this modeling paradigm.
\end{remark}

The governing equations of the classical water wave problem are the following, \cite{Lamb1932, Stoker1958, Mei1994, Whitham1999}:
\begin{eqnarray}
  \Delta\phi = \grad^2\phi + \partial^2_{zz}\phi &=& 0, 
  \qquad (\x, z) \in \Omega\times [-h, \eta], \label{eq:laplace} \\
  \dt\eta + \grad\phi\cdot\grad\eta - \dz\phi &=& 0, 
  \qquad z = \eta(\x, t), \label{eq:kinematic} \\
  \dt\phi + \half|\grad\phi|^2 + \half(\dz\phi)^2 + g\eta &=& 0, 
  \qquad z = \eta(\x,t), \label{eq:bernoulli} \\
  \dt h + \grad\phi\cdot\grad h + \dz\phi &=& 0, 
  \qquad z = -h(\x,t), \label{eq:bottomkin}
\end{eqnarray}
where $\phi$ is the velocity potential, $g$ the acceleration due to gravity force and $\grad = (\partial_x, \partial_y)$ denotes the gradient operator in horizontal Cartesian coordinates. The fluid incompressibility and flow irrotationality assumptions lead to the Laplace equation (\ref{eq:laplace}) for the velocity potential $\phi(\x, z, t)$. 

The main difficulty of the water wave problem lies on the boundary conditions. Equations (\ref{eq:kinematic}) and (\ref{eq:bottomkin}) express the free surface and bottom impermeability, while the Bernoulli condition (\ref{eq:bernoulli}) expresses the free surface isobarity respectively.

Function $h(\x,t)$ represents the ocean's bathymetry (depth below the still water level, see Figure~\ref{fig:sketch}) and is assumed to be known. The dependence on time is included in order to take into account the bottom motion during an underwater earthquake \cite{Dias2006, Dutykh2006, ddk, Kervella2007, Dutykh2007a}.

For the exposition below we will need also to compute unitary exterior normals to the fluid domain. The normals at the free surface $S_f$ and at the bottom $S_b$ are given by the following expressions respectively:
\begin{equation*}
  \n_f = \frac{1}{\sqrt{1 + |\grad\eta|^2}}\left|
    \begin{array}{c}
      -\grad\eta \\
      1
    \end{array}
  \right.,
  \qquad
  \n_b = \frac{1}{\sqrt{1 + |\grad h|^2}}\left|
    \begin{array}{c}
      -\grad h \\
      -1
    \end{array}
  \right. .
\end{equation*}

In 1968 V.~Zakharov proposed a different formulation of the water wave problem based on the trace of the velocity potential at the free surface \cite{Zakharov1968}:
\begin{equation*}
  \phis(\x, t) := \phi(\x, \eta(\x,t), t).
\end{equation*}
This variable plays a role of the generalized momentum in the Hamiltonian description of water waves \cite{Zakharov1968, Dias2006a}. The second canonical variable is the free surface elevation $\eta$.

Another important ingredient is the normal velocity at the free surface $v_n$ which is defined as:
\begin{equation}\label{eq:normalv}
  v_n (\x,t) := \sqrt{1 + |\grad\eta|^2}\left.\pd{\phi}{\n_f}\right|_{z=\eta} = 
  \left.(\dz\phi - \grad\phi\cdot\grad\eta)\right|_{z=\eta}.
\end{equation}
Kinematic and dynamic boundary conditions \eqref{eq:kinematic}, \eqref{eq:bernoulli} at the free surface can be rewritten in terms of $\phis$, $v_n$ and $\eta$ \cite{Craig1992, Craig1993, Fructus2005}:
\begin{equation}\label{eq:dynamics}
 \begin{array}{rl}
  \dt\eta - \D_\eta(\phis) &= 0, \\
  \dt\phis + \half|\grad\phis|^2 + g\eta
  - \frac{1}{2(1+|\grad\eta|^2)}\bigl[\D_\eta(\phis) + 
   \grad\phis\cdot\grad\eta \bigr]^2 &= 0.
 \end{array}
\end{equation}
Here we introduced the so-called Dirichlet-to-Neumann operator (D2N) \cite{Coifman1985, Craig1993} which maps the velocity potential at the free surface $\phis$ to the normal velocity $v_n$:
\begin{eqnarray*}
  \D_\eta: & \phis \mapsto v_n = \sqrt{1 + |\grad\eta|^2}
  \left.\pd{\phi}{\n_f}\right|_{z=\eta} \\
  & \left|
  \begin{array}{rl}
    \grad^2\phi + \partial^2_{zz}\phi &= 0, \quad (\x, z) \in \Omega\times [-h, \eta], \\
    \phi &= \phis, \quad z=\eta, \\
    \sqrt{1 + |\grad h|^2}\displaystyle{\pd{\phi}{\n_b}} &= \dt h, \quad z=-h.
  \end{array}\right.
\end{eqnarray*}
The name of this operator comes from the fact that it makes a correspondance between Dirichlet data $\phis$ and Neumann data $\sqrt{1+|\grad\eta|^2}\left. \displaystyle{\pd{\phi}{\n_f}} \right|_{z=\eta}$ at the free surface.

So, the water wave problem can be reduced to a system of two PDEs \eqref{eq:dynamics} governing the evolution of the canonical variables $\eta$ and $\phis$. For the tsunami generation problem we approximate and we compute efficiently the D2N map $\D_\eta(\phis)$  using the Weakly Nonlinear (WN) model described in \cite{Dutykh2010a}. This relies on the approximate solution of the 3D Laplace equation in a perturbed strip-like domain using the Fourier transform ($\hat\phis := \F[\phis]$, $\eta = \F^{-1}[\hat\eta]$):
\begin{equation*}
  \hat\D_\eta(\phis) = \hat{\phis}|\k|\tanh(|\k|H) + \hat{f}\sech(|\k|H) - \F\Bigl[\F^{-1}\bigl[i\k\hat{\phis}\bigr]\cdot\F^{-1}\bigl[i\k\hat{\eta}\bigr]\Bigr],
\end{equation*}
where $\k$ is the wavenumber and $f$ is the bathymetry forcing term 
\begin{equation*}
  f(\x,t) := -\dt h - \left.\grad\phi\right|_{z=-h}\cdot\grad h.
\end{equation*}
Several details on the time integration procedure can be also found in \cite{Dutykh2010a}. The resulting method is only weakly nonlinear and analogous at some point to the first order approximation model proposed in \cite{Guyenne2007}. As the hydrodynamical models, Tanioka and Satake (1996) \cite{Tanioka1996} used the Linear Shallow Water equations \cite{Satake1995}, while Song \emph{et al.} (2008) employed a hydrostatic ocean circulation model \cite{Song2006}.

Recently, it was shown that a tsunami generation process is essentially linear, \cite{Kervella2007, Saito2009}. However, our WN approach will take into account some nonlinear effects when they become important, for example when rapid changes in the bathymetry are present. This is possible when the generation region involves a wide range of depths from deep to shallow regions (see Figures \ref{fig:bathy1} and \ref{fig:bathy2}).

\subsubsection{Tsunami wave energy}\label{sec:energy}

In this study we are more particularly interested in the evolution of the generated wave energy \cite{Dutykh2009b}. In the case of free surface incompressible flows, the kinetic and potential energies, denoted by $K$ and $\Pi$ respectively, are completely determined by the velocity field and the free surface elevation:
\begin{equation*}
  K(t) := \frac{\rho}{2}\int\limits_{-h}^{\eta}\iint\limits_{\Omega}|\grad\phi|^2\; d\x\,dz, \qquad
  \Pi(t) := \frac{\rho g}{2}\iint\limits_{\Omega}\eta^2\;d\x.
\end{equation*}
The definition of the kinetic energy $K(t)$ involves an integral over the three dimensional physical domain $\Omega\times[-h,\eta]$. We can reduce the integral dimension using the fact that the velocity potential $\phi$ is a harmonic function:
\begin{equation*}
  |\grad\phi|^2 = \div(\phi\grad\phi) - \phi\underbrace{\Delta\phi}_{=0} \equiv \div(\phi\grad\phi).
\end{equation*}
Consequently, the kinetic energy can be rewritten as follows:
\begin{equation*}
  K(t) = \frac{\rho}{2}\iint\limits_{S_f+S_b} \phi\grad\phi\cdot\n\;d\sigma = 
  \underbrace{\frac{\rho}{2}\iint\limits_{\Omega} \phis\D_\eta(\phis) \;d\x}_{(I)}
  + \underbrace{\frac{\rho}{2}\iint\limits_{\Omega}\check{\phi}\dt h\;d\x}_{(II)},
\end{equation*}
where $\check{\phi}$ denotes the trace of the velocity potential at the bottom $\left.\phi\right|_{z=-h}$ (see D.~Clamond \& D.~Dutykh (2012), \cite{Clamond2009}). In order to obtain the last equality we used the free surface and the bottom kinematic boundary conditions \eqref{eq:bernoulli}, \eqref{eq:bottomkin}. The first integral (I) is classical and represents the change of kinetic energy under the free surface motion while the second one (II) is the forcing term due to the bottom deformation. The total energy\footnote{We note that the total energy is not conserved during the tsunami generation phase due to the forcing term (II) coming from the bottom kinematic boundary condition.} is defined as the sum of kinetic and potential ones:
\begin{equation*}
  E(t) := K(t) + \Pi(t) = \frac{\rho}{2}\iint\limits_{\Omega} \phis\D_\eta(\phis) \;d\x + \frac{\rho}{2}\iint\limits_{\Omega}\check{\phi}\dt h\;d\x + \frac{\rho g}{2}\iint\limits_{\Omega}\eta^2\;d\x.
\end{equation*}

Below we will compute the evolution of the kinetic, potential and total energies beneath moving bottom.

\section{Numerical results}\label{sec:numres}

\begin{figure}
  \centering
  \includegraphics[width=0.8\textwidth]{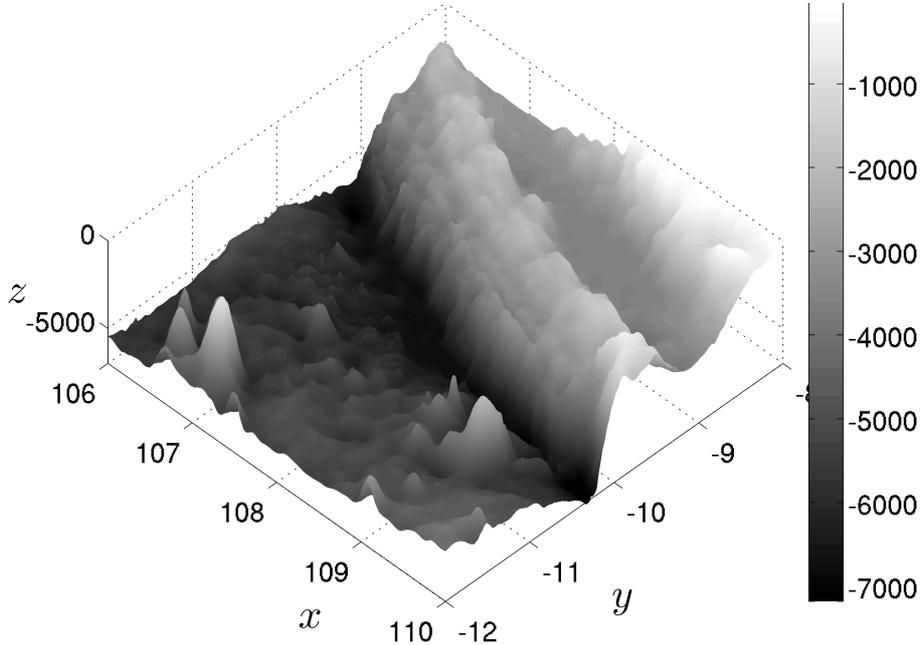}
  \caption{Side view of the bathymetry, cf. also Figure \ref{fig:bathy1}.}
  \label{fig:bathy2}
\end{figure}

The proposed approach will be directly illustrated on the Java 2006 event. The July 17, 2006 Java earthquake involved thrust faulting in the Java's trench and generated a tsunami wave that inundated the southern coast of Java, \cite{Ammon2006, Fritz2007}. The estimates of the size of the earthquake, \cite{Ammon2006}, indicate a seismic moment of $6.7 \times 10^{20}$ $N\cdot m$, which corresponds to the magnitude $M_w = 7.8$. Later this estimation was refined to $M_w = 7.7$, \cite{Ji2006}. Like other events in this region, Java's event had an unusually low rupture speed of $1.0$ -- $1.5$ km/s (we take the value of $1.1$ km/s according to the finite fault solution \cite{Ji2006}), and occurred near the up-dip edge of the subduction zone thrust fault. According to C.~Ammon {\em et al}, \cite{Ammon2006}, most aftershocks involved normal faulting. The rupture propagated approximately $200$ km along the trench with an overall duration of approximately $185$ s. The fault's surface projection along with ocean ETOPO1 bathymetric map are shown in Figures \ref{fig:bathy1} and \ref{fig:bathy2}. We note that Indian Ocean's depth of the region considered in this study varies between 7186 and 20 meters in the shallowest regions which may imply local importance of nonlinear effects.

\begin{remark}
We have to mention that the finite fault inversion for this earthquake was also performed by the Caltech team, \cite{Ozgun2006}. The main differences with the USGS inversion consist on the employed dataset. To our knowledge, A.~Ozgun~Konca and his collaborators include also displacements measured with GPS-based techniques. Consequently, they came to the conclusion that the July 17, Southern Java earthquake magnitude was $M_w = 7.9$. The energy of a tsunami wave generated according to this solution will be discussed below.
\end{remark}

\begin{table}
\begin{center}
\begin{tabular}{c||c}
  \hline\hline
  Ocean water density, $\rho$, $kg/m^3$ & 1027.0 \\
  \hline
  Gravity acceleration, $g$, $m/s^2$ & 9.81 \\
  \hline
  Epicenter location (Lon, Lat) & (107.345$^\circ$, -9.295$^\circ$) \\
  \hline
  Rupture velocity, $v_r$, $km/s$ & 1.1 \\
  \hline
  Rise time, $t_0$, $s$ & 8.0 \\
  \hline
  Number of Fourier modes in $x$ & 256 \\
  \hline
  Number of Fourier modes in $y$ & 256 \\
  \hline\hline
\end{tabular}
\caption{Values of physical and numerical parameters used in simulations.}
\label{tab:fluid}
\end{center}
\end{table}

The numerical solutions presented below are given by the Weakly Nonlinear (WN) model. A uniform grid of $256 \times 256$ points\footnote{Since we use a pseudo-spectral method, the convergence is expected to be exponential and this number of harmonics should be sufficient to capture all scales important for phenomena that we consider here.} is used in all computations below. The time step $\Delta t$ is chosen adaptively according to the RK(4,5) method proposed in \cite{Dormand1980}. The problem is integrated numerically during $T = 255$ s which is a sufficient time interval for the bottom to take its final shape ($< 220$ s) and of the resulting tsunami wave to enter into the propagation stage. The values of various physical and numerical parameters used in simulations are given in Table \ref{tab:fluid}.

\begin{figure}
  \centering
  \includegraphics[width=0.65\textwidth]{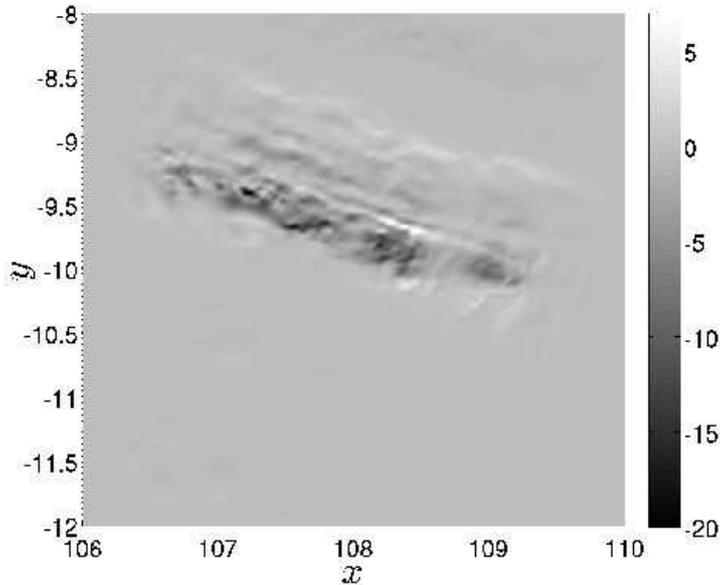}
  \caption{Contribution of the horizontal displacements field into the sea bed deformation expressed in percentage of the maximum vertical displacement. The grey color corresponds to negligible values while white and black zones show the most important contributions. The colorbar shows the percentage of the signed difference going from -21\% to +7\%.}
  \label{fig:hordisp}
\end{figure}

We begin our numerical investigations by quantifying the contribution of horizontal displacements into the sea bed deformation process. For this purpose we consider the difference $d_h(\x)$ between the deformed bottom under the action of only horizontal displacements in their steady state ($t\to +\infty$) and the initial configuration:
\begin{equation*}
  d_h (\x) := \frac{h_0\bigl(\x - \left.\u_{1,2}(\x, t)\right|_{t\to +\infty}\bigr) - h_0(\x)}{\max\limits_{\x, t\to+\infty}|u_3(\x, t)|}.
\end{equation*}
In Figure \ref{fig:hordisp} we present the quantitative effect of horizontal displacements relative to the maximum vertical displacement $\max\limits_{\x}|u_3(\x, t=+\infty)|$. The computations we performed show that the maximum amplitude of the bottom variation due to the action of horizontal displacements reaches 21\% of the maximum amplitude of the vertical displacement. In practice it means that locally (depending on the bathymetry shape and the slip distribution) we cannot completely neglect the effect of horizontal motion.

\begin{figure}
  \centering
  \includegraphics[width=0.6\textwidth]{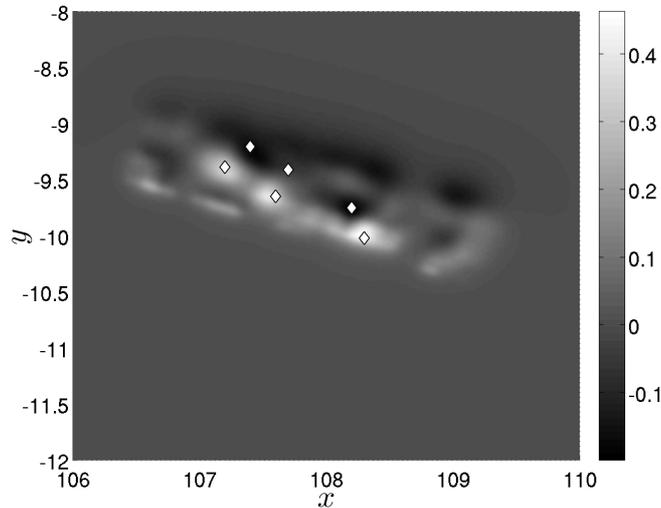}
  \caption{Location of the six numerical wave gauges (indicated by the symbol $\diamond$) superposed with the steady state coseismic bottom displacement (only the vertical component in meters is represented here).}
  \label{fig:gauges}
\end{figure}

\begin{figure}
  \centering
  \subfigure[Gauge at $(107.2^\circ, -9.388^\circ)$]%
  {\includegraphics[width=0.495\textwidth]{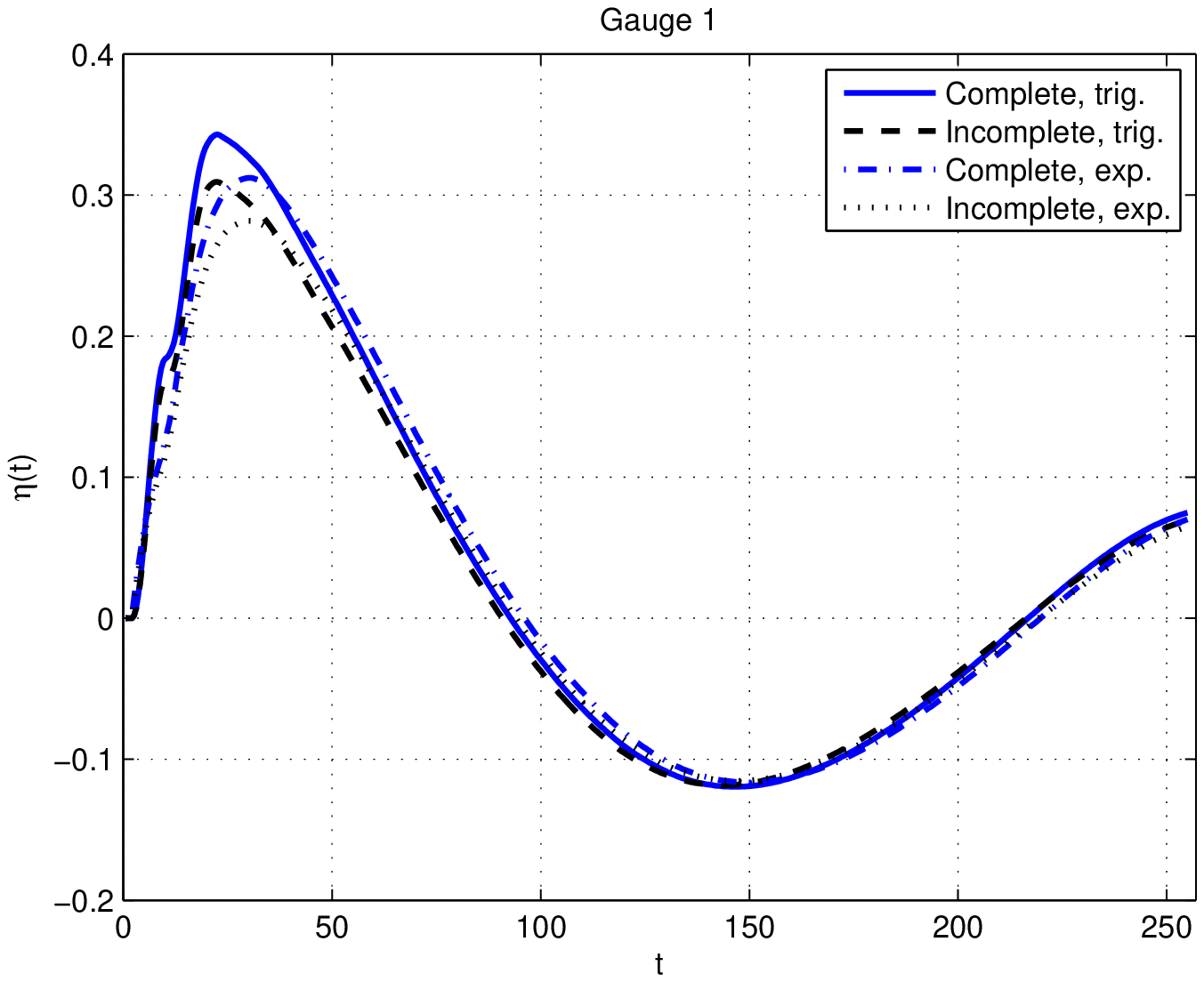}}
  \subfigure[Gauge at $(107.4^\circ, -9.205^\circ)$]%
  {\includegraphics[width=0.495\textwidth]{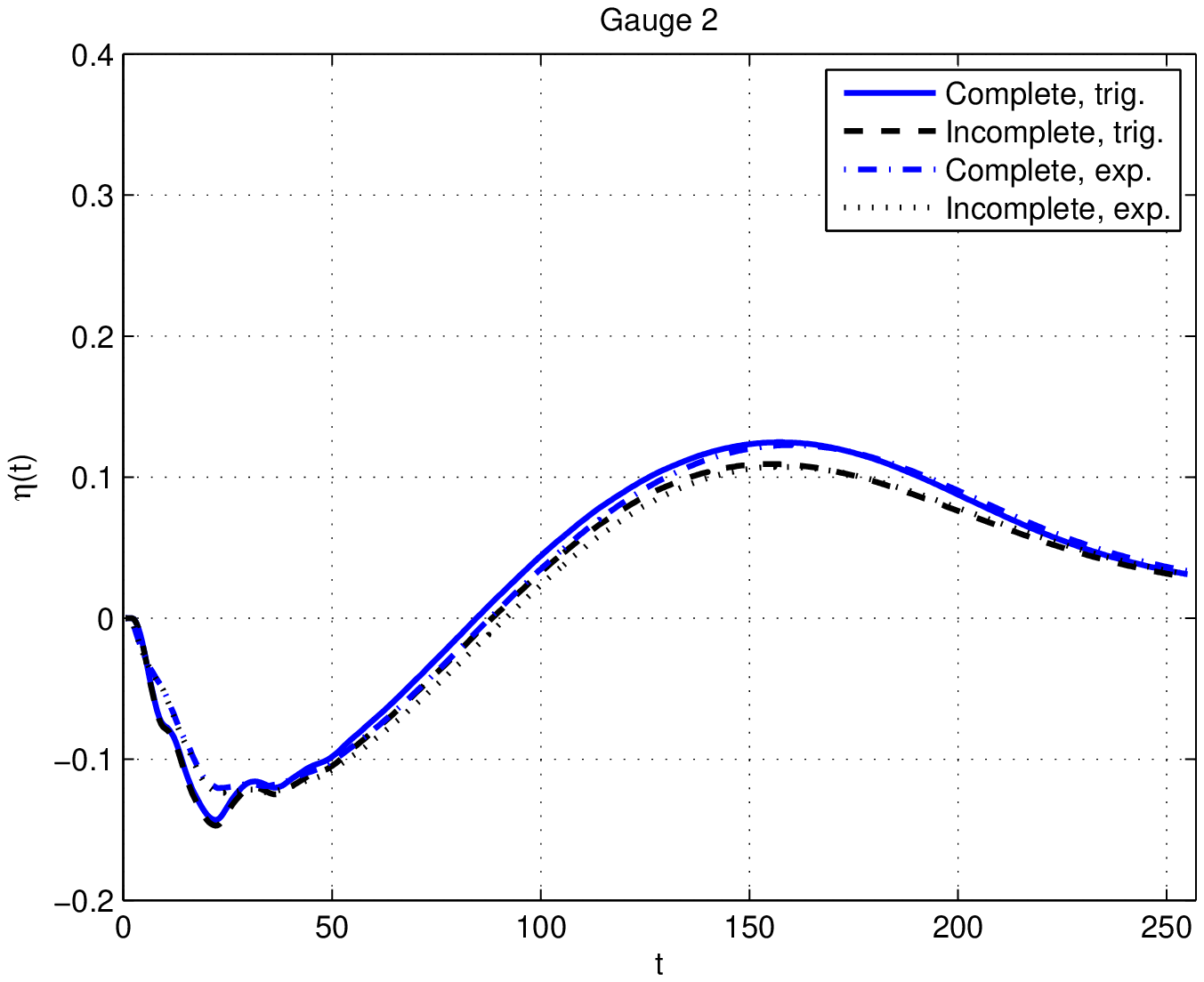}}
  \subfigure[Gauge at $(107.6^\circ, -9.648^\circ)$]%
  {\includegraphics[width=0.495\textwidth]{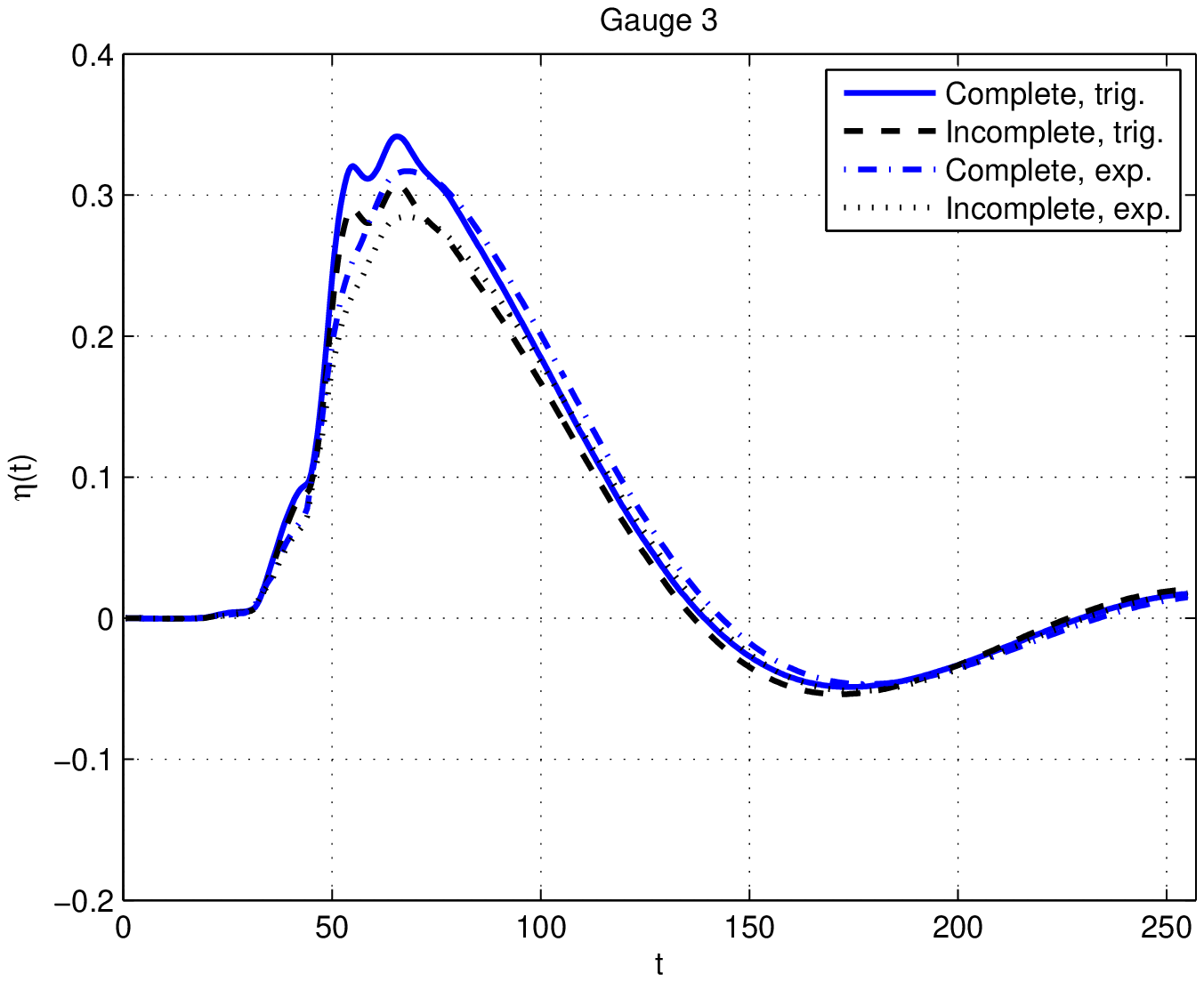}}
  \subfigure[Gauge at $(107.7^\circ, -9.411^\circ)$]%
  {\includegraphics[width=0.495\textwidth]{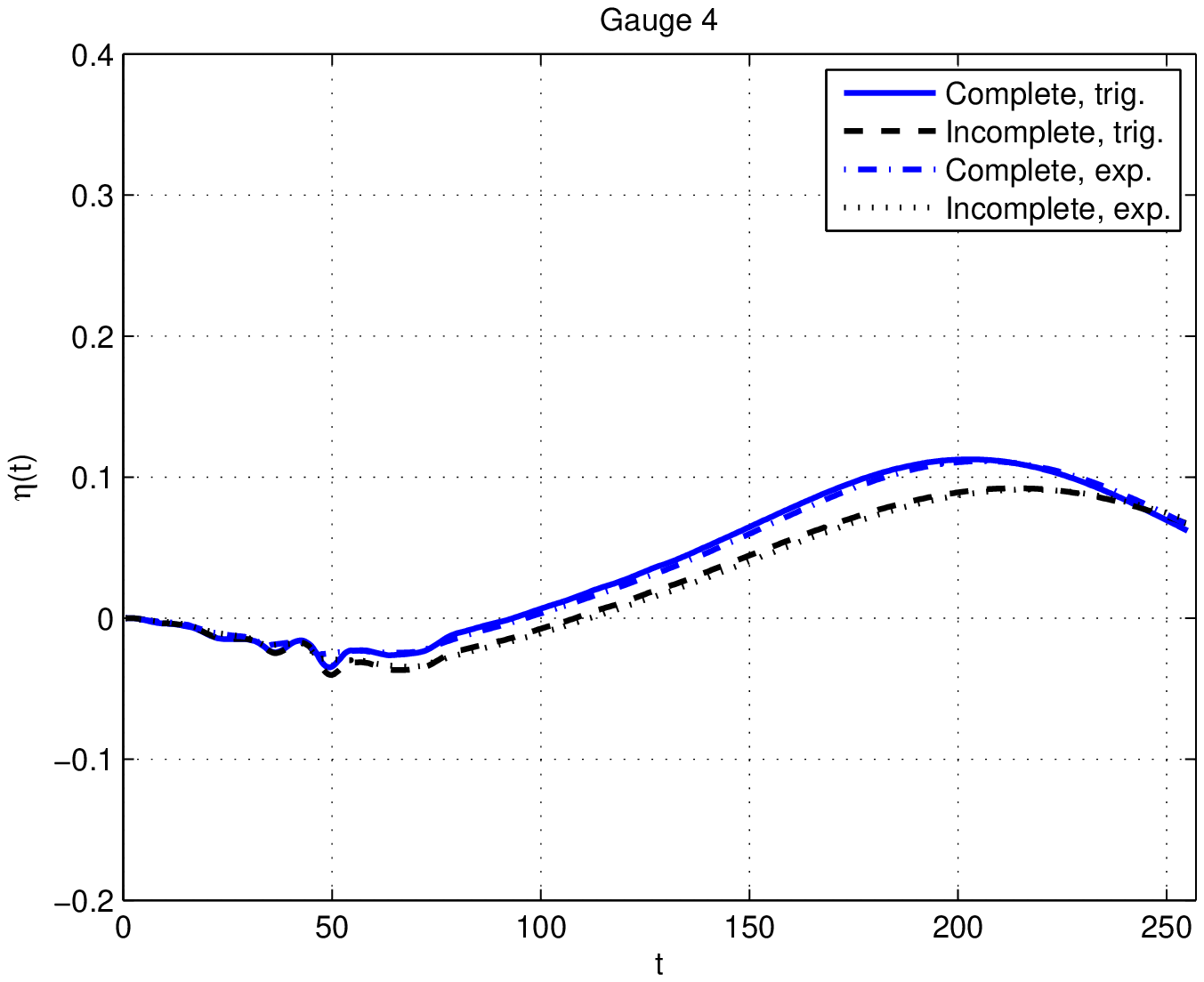}}
  \subfigure[Gauge at $(108.3^\circ, -10.02^\circ)$]%
  {\includegraphics[width=0.495\textwidth]{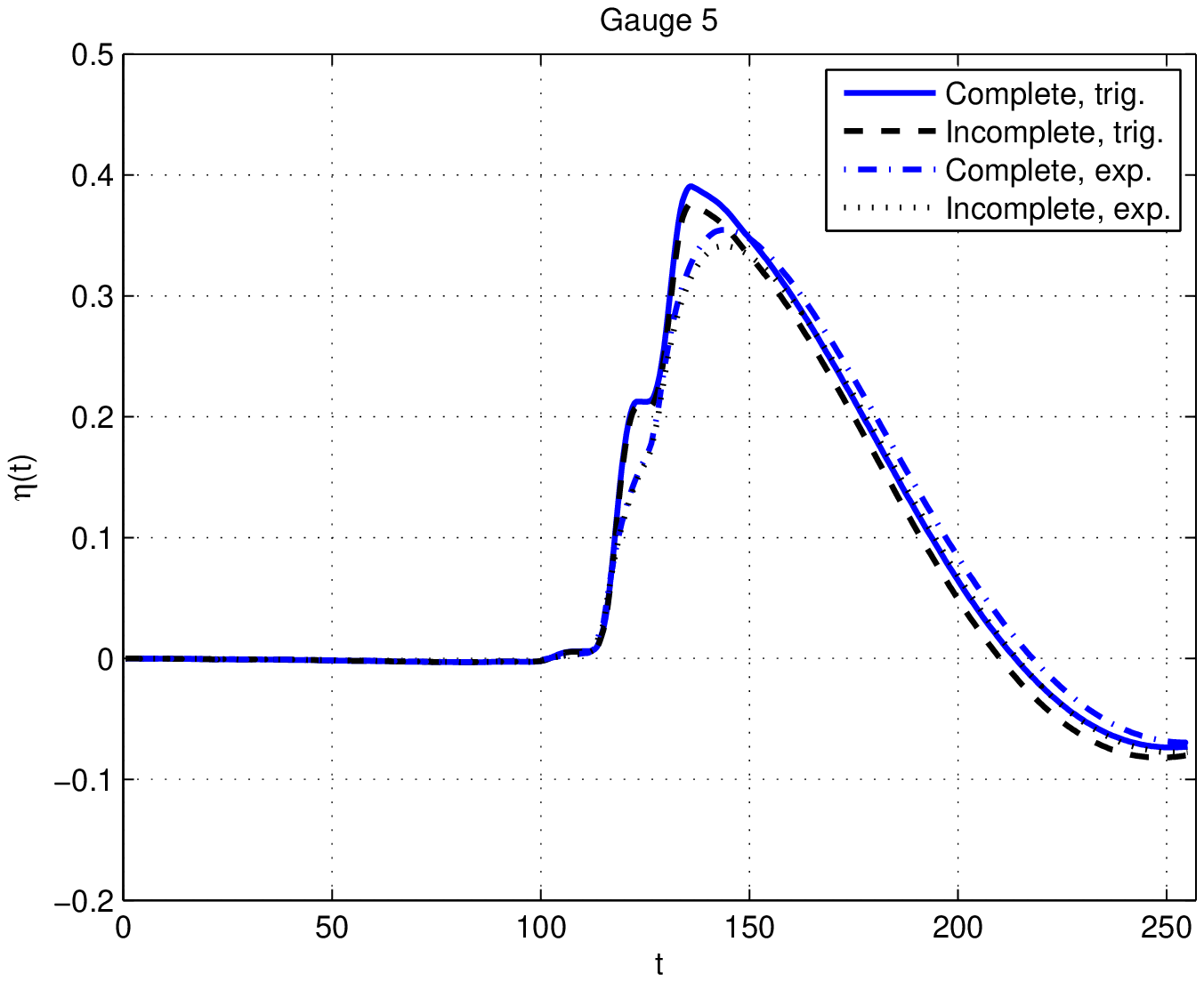}}
  \subfigure[Gauge at $(108.2^\circ, -9.75^\circ)$]%
  {\includegraphics[width=0.495\textwidth]{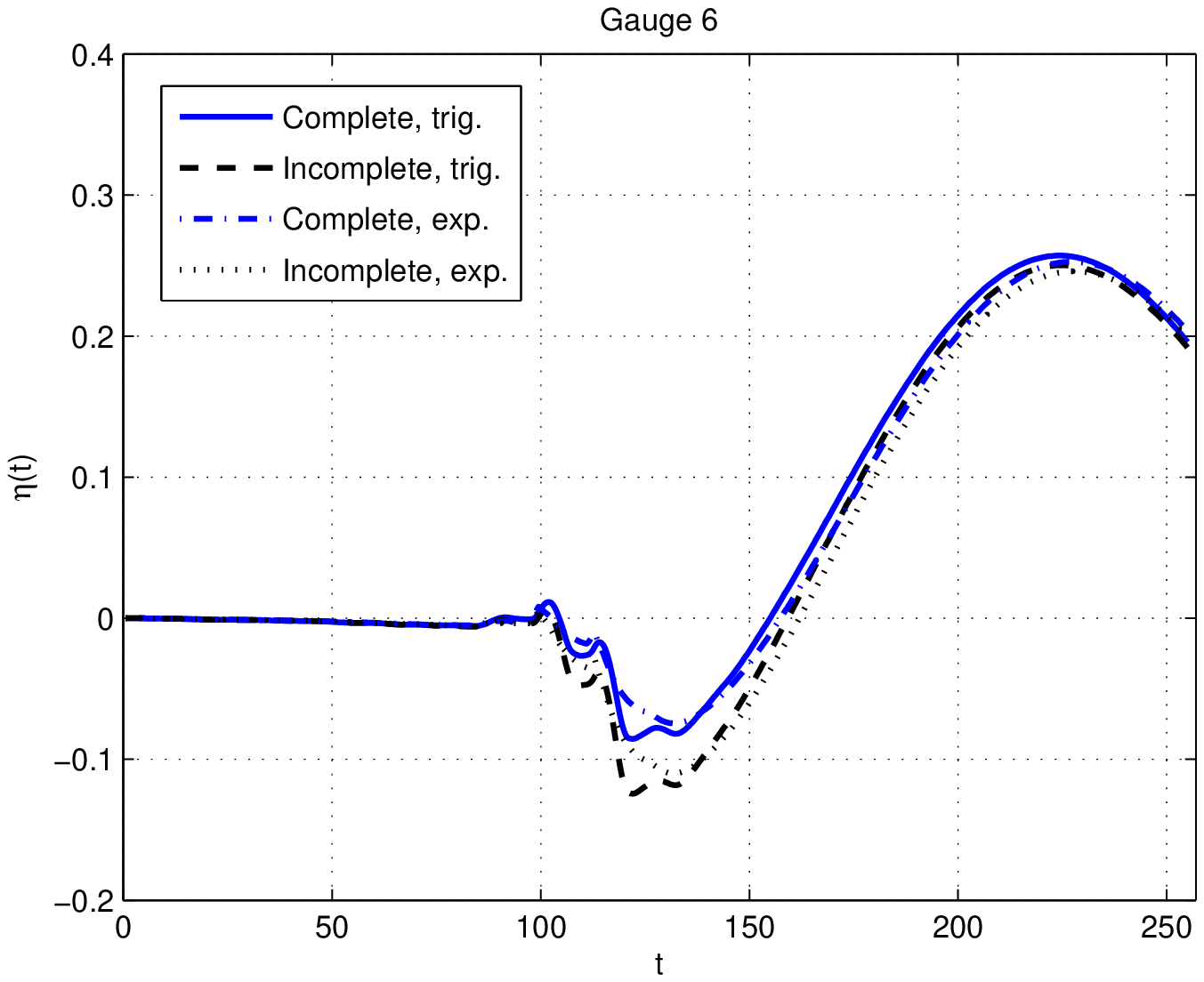}}
  \caption{Free surface elevations computed numerically at six wave gauges located approximately in local extrema of the static bottom displacement. The vertical axis is represented in meters and time is given in seconds. The black lines correspond to the wave generated only by the vertical displacements (incomplete generation) while blue lines take also into account the horizontal displacements contribution (complete scenario).}
  \label{fig:gaugedata}
\end{figure}

The next step consists in quantifying the impact of horizontal displacements onto free surface motion. We put six numerical wave gauges at the following locations: (a) $(107.2^\circ$, $-9.388^\circ)$, (b) $(107.4^\circ$, $-9.205^\circ)$, (c) $(107.6^\circ$, $-9.648^\circ)$, (d) $(107.7^\circ$, $-9.411^\circ)$, (e) $(108.3^\circ$, $-10.02^\circ)$, (f) $(108.2^\circ$, $-9.75^\circ)$. The locations of the wave gauges are represented on Figure \ref{fig:gauges} along with the static sea bed displacement. Wave gauges are intentionally put in places where the largest waves are expected. Synthetic wave gauge records are presented in Figure \ref{fig:gaugedata}. We consider the following four scenarios:
\begin{itemize}
  \item trigonometric, complete (blue solid line)
  \item trigonometric, incomplete (black dashed line)
  \item exponential, complete (blue dash-dotted line)
  \item exponential, incomplete (black dotted line)
\end{itemize}
The importance of the nonlinear effects have already been addressed in several previous studies, \cite{Kervella2007, Saito2009}. In the framework of the Weakly Nonlinear approach we studied this question in a recent companion paper \cite{Dutykh2010a} using only the vertical deformation. A fairly good agreement has been observed with the Cauchy-Poisson (linear, fully dispersive) formulation in accordance with preceding results, cf. \cite{Kervella2007, Saito2009}. Consequently, in the present study we decided to focus on the further comparison between complete/incomplete approaches and exponential/trigonometric scenarios, \cite{Hammack, ddk, Mitsotakis2007}. These results are discussed hereafter.

\begin{figure}
  \centering
  \includegraphics[width=0.65\textwidth]{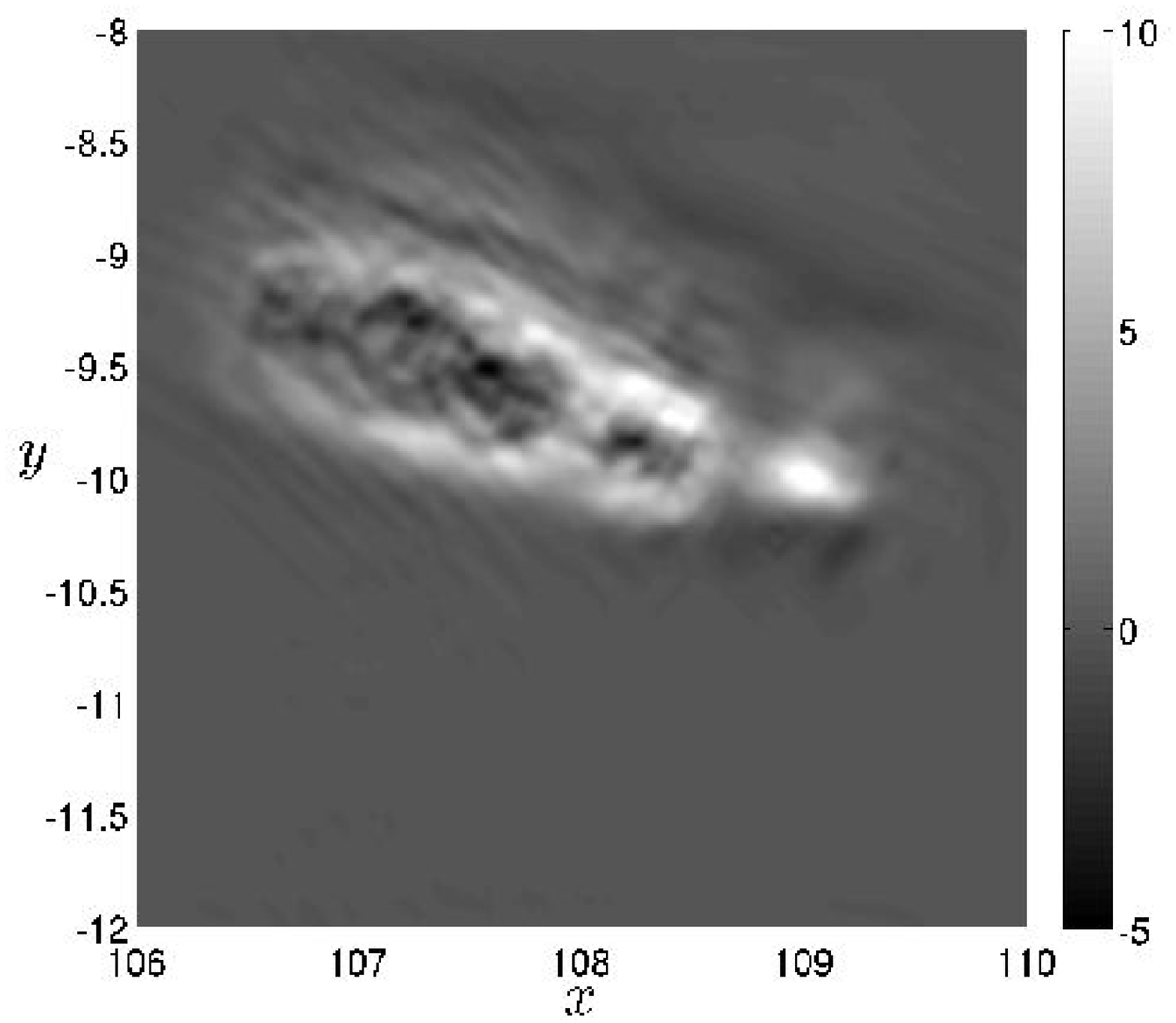}
  \caption{Relative difference between the free surface elevation at $t_e=220$ seconds computed according to the complete $\eta_c(\x, t_e)$ (with horizontal displacements) and incomplete $\eta_i(\x, t_e)$ (only vertical component) tsunami generation scenarios. The bottom moves according to the trigonometric scenario \eqref{eq:expscenario}. The vertical scale is given in percents of the maximum amplitude of the incomplete scenario --- $\max\limits_{\x}\eta_i(\x, t_e)$.}
  \label{fig:diffHV}
\end{figure}

We note that the trigonometric scenario leads in general to slightly larger amplitudes than the exponential bottom motion. It is not surprising since the exponential scenario prescribes smoother and less rapid change in the bottom for the same rise time parameter value (see Figure \ref{fig:scenario}). Later we will consider the trigonometric scenario unless otherwise noted.

Then, it can be seen that in most cases horizontal displacements lead to an increase in the wave amplitude but not always as it can be observed in Figure \ref{fig:gaugedata}(f) (the last wave gauge located at $(108.2^\circ, -9.75^\circ)$). We investigated more thoroughly this question. Figure \ref{fig:diffHV} shows the relative difference between free surface elevations computed according to complete $\eta_c(\x, t_e)$ and incomplete $\eta_i(\x, t_e)$ scenarios. More precisely, we plot the following quantity (for the trigonometric scenario):
\begin{equation*}
  d(\x) := \frac{\eta_c(\x,t_e) - \eta_i(\x,t_e)}{\max\limits_{\x}|\eta_i(\x,t_e)|}, \qquad t_e = 220\; s.
\end{equation*}
Time $t_e = 220$ s has been chosen because at that moment the bottom has been stabilized and the waves enter into the free propagation regime. In Figure \ref{fig:diffHV} the grey color corresponds to the zero value of the difference $d(\x)$, while the white color shows regions where the wave is amplified by horizontal displacements by approximately $10\%$. On the contrary, black zones show an attenuation effect of horizontal sea bed motion (about $-5\%$). Recall that all values are given in terms of the maximum amplitude $\max\limits_{\x}|\eta_i(\x,t_e)|$ percentage of the incomplete generation approach. Some connection with the results presented in Figure \ref{fig:hordisp} can be noticed.

\begin{figure}
  \centering
  \subfigure[Potential energy $\Pi(t)$, J]%
  {\includegraphics[width=0.495\textwidth]{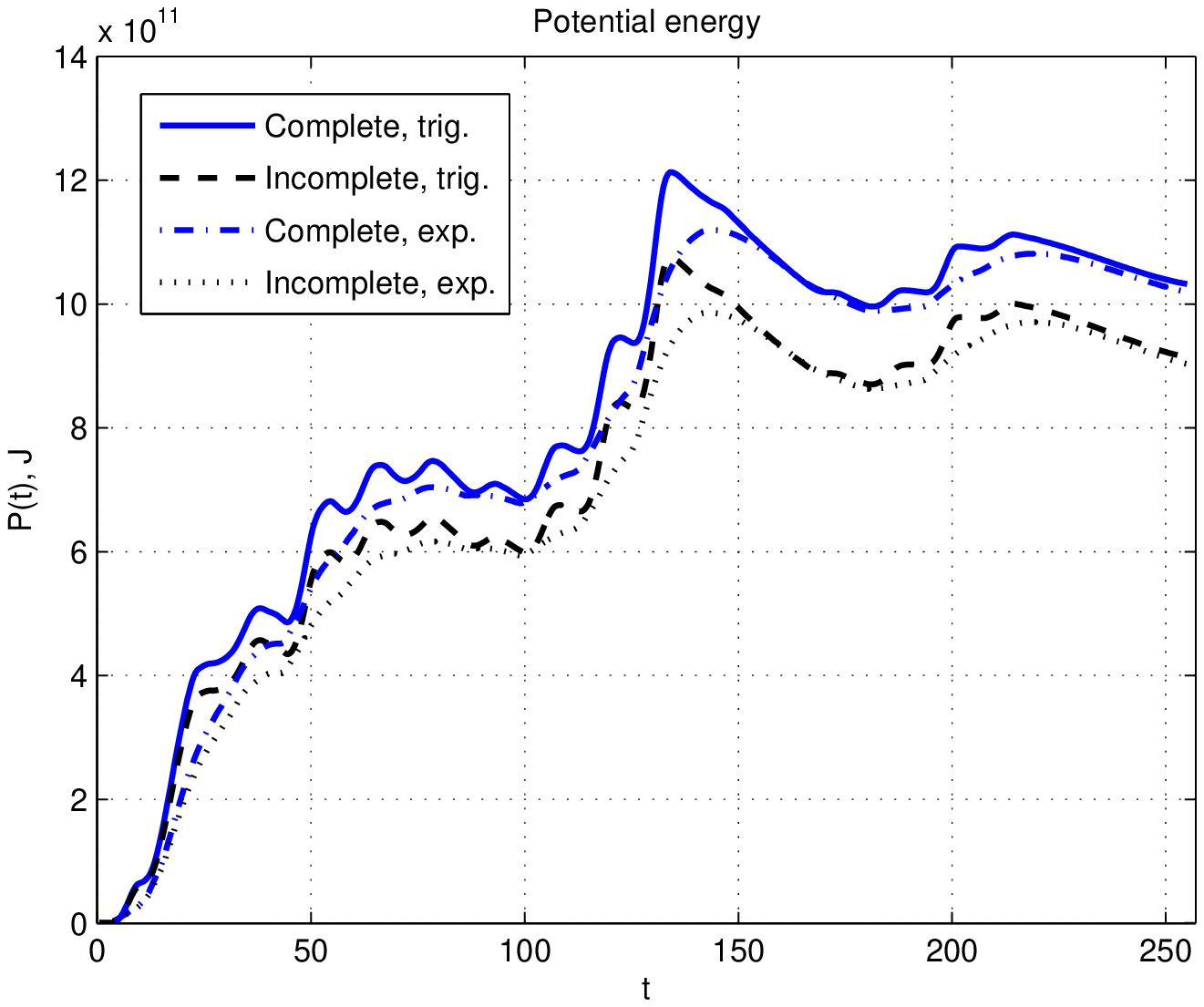}}
  \subfigure[Kinetic energy $K(t)$, J]%
  {\includegraphics[width=0.495\textwidth]{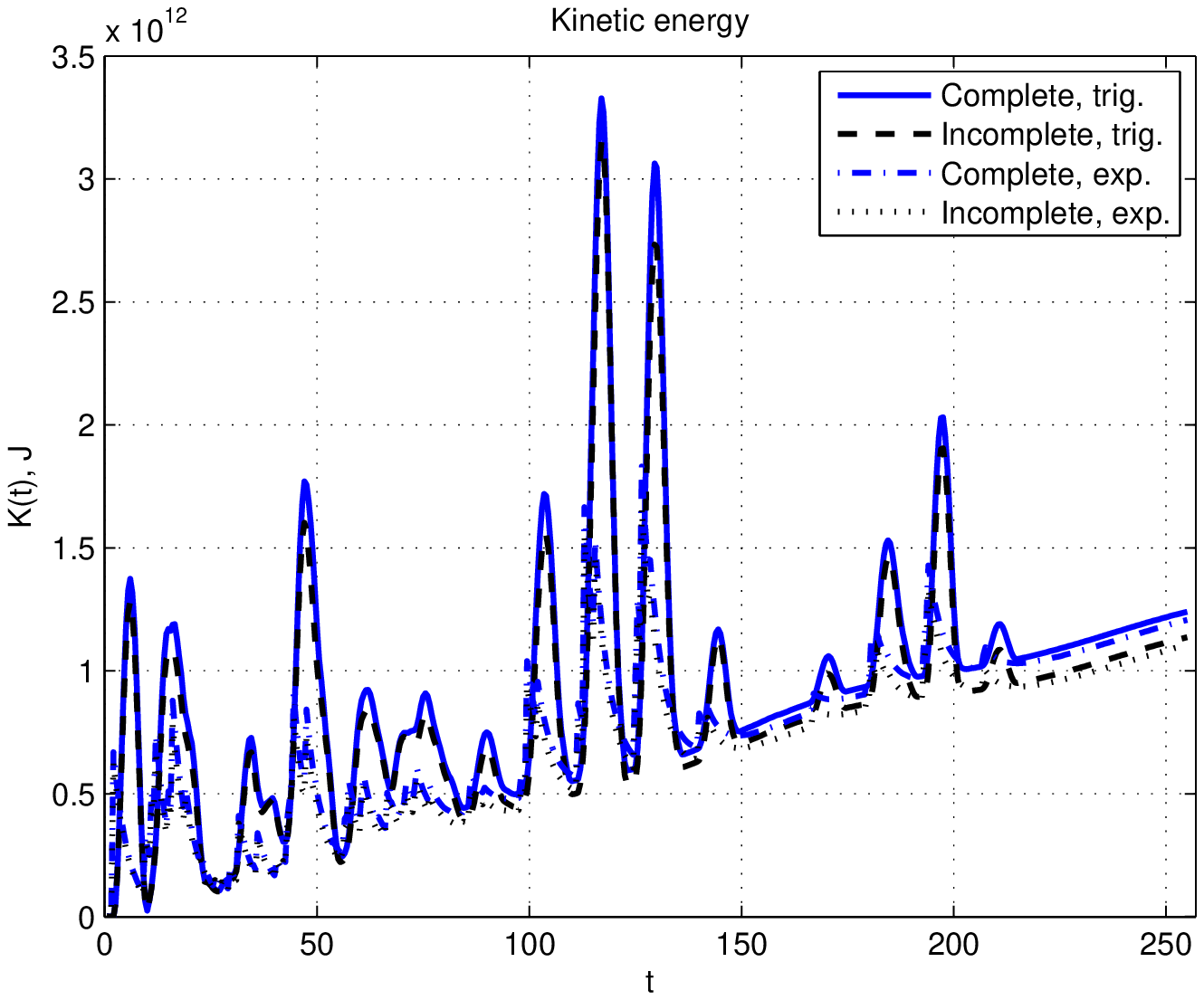}}
  \caption{Energy evolution during our simulations in the complete (blue solid line) and incomplete (black dotted line) scenarios. Note that scales are different on the left and right images. The time $t$ is given in seconds.}
  \label{fig:energies}
\end{figure}

Finally, we study the evolution of kinetic, potential and total energies during the tsunami generation process described in Section \ref{sec:energy}. Specifically we are interested in quantifying the contribution of horizontal displacements into tsunami energy balance. Figure \ref{fig:energies} shows the evolution of potential (\ref{fig:energies}(a)) and kinetic (\ref{fig:energies}(b)) energies for four cases already mentioned above. Here again blue lines refer to complete generation scenarios while black lines -- to vertical displacements only. Consecutive peaks in the kinetic energy come from the activation of new fault segments in accordance with the rupture propagation along the fault. The energy curves vary in an analogous way with the tide gauges records (see Figure \ref{fig:gaugedata}). Namely, the trigonometric scenario leads to slightly higher energies than the exponential one. However, once the bottom motion stops, this difference becomes negligible. The inclusion of horizontal displacements has a much more visible effect with higher energies.

\begin{figure}
  \centering
  \subfigure[Potential energy $\Pi(t)$, J]%
  {\includegraphics[width=0.495\textwidth]{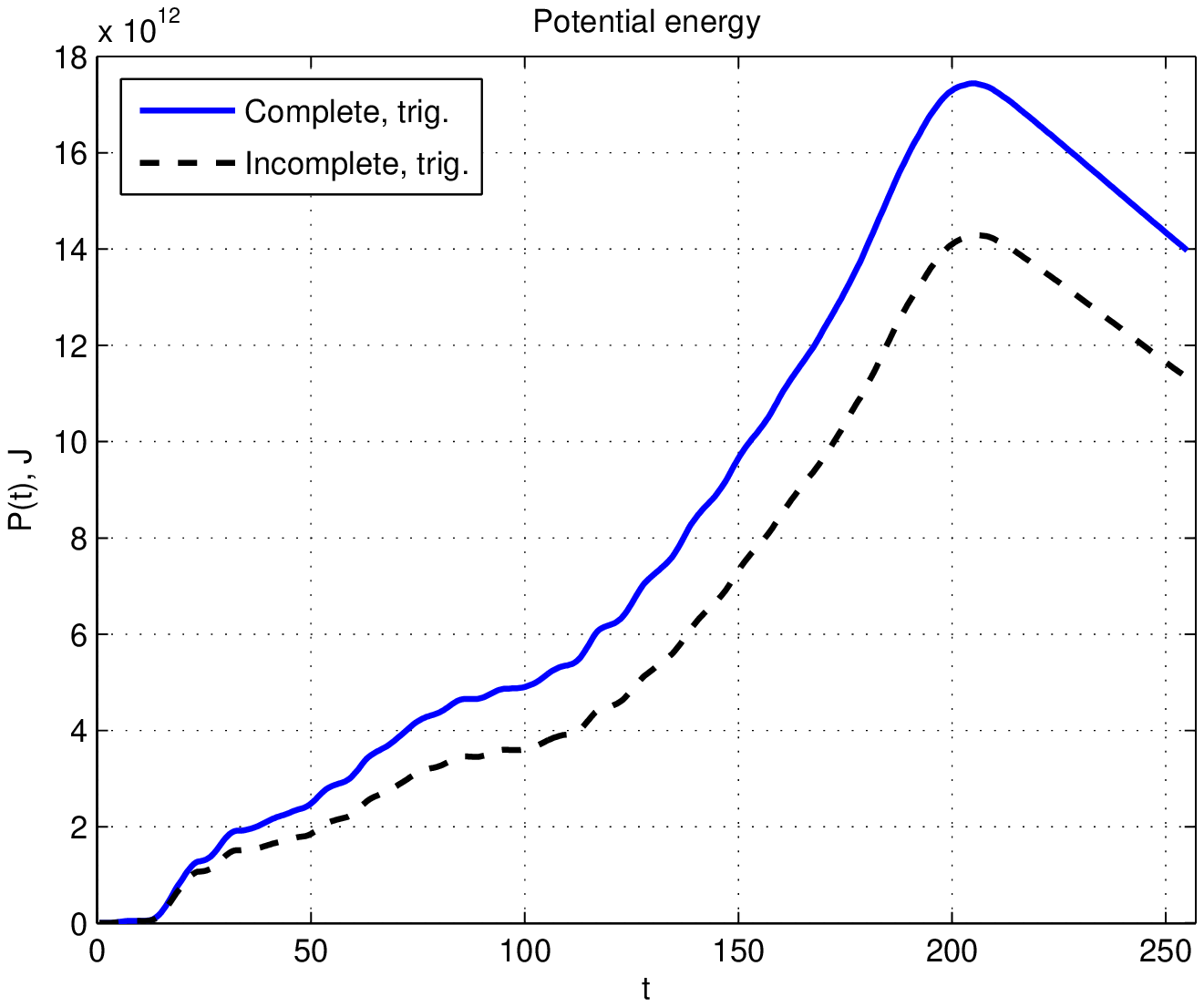}}
  \subfigure[Kinetic energy $K(t)$, J]%
  {\includegraphics[width=0.495\textwidth]{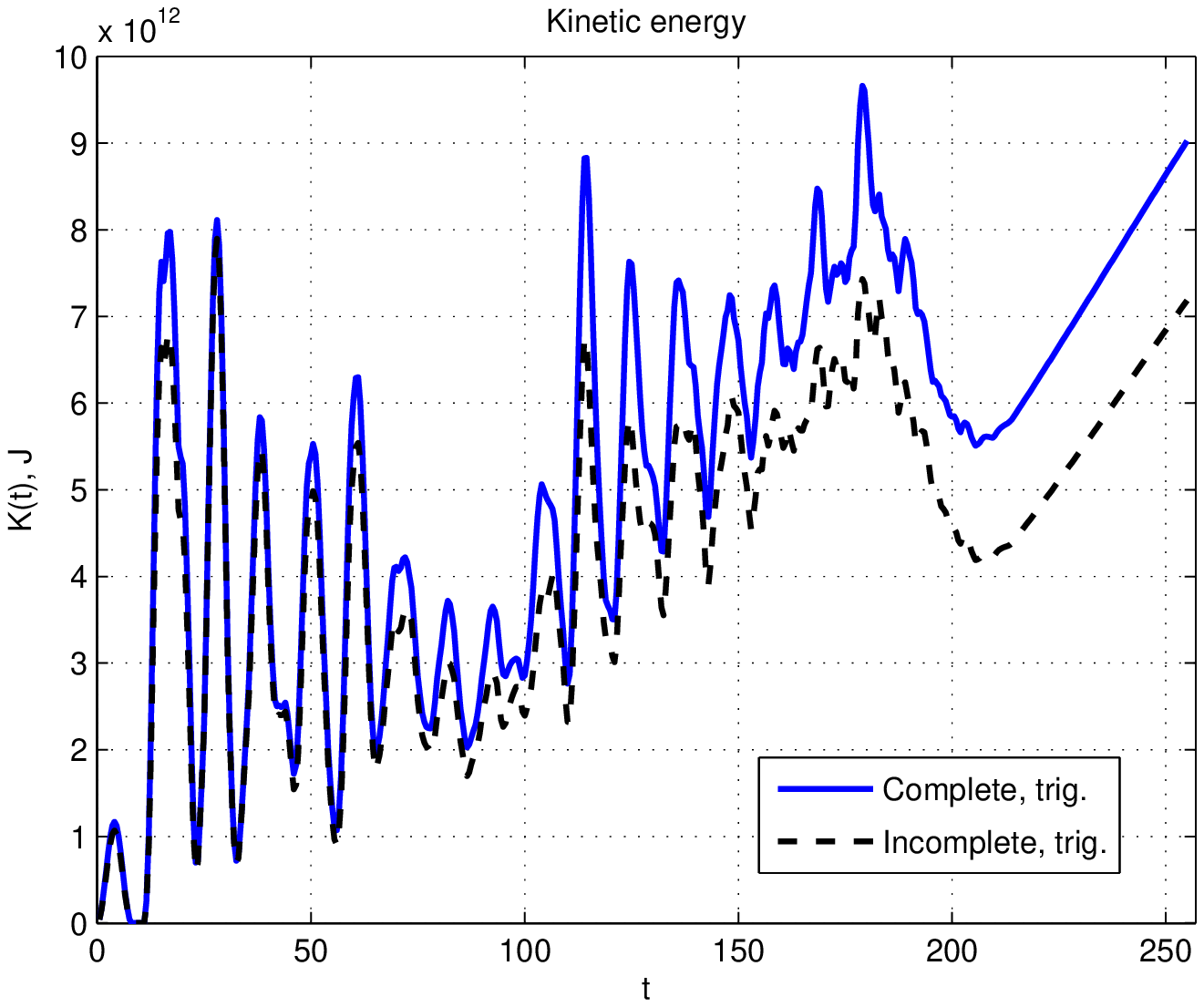}}
  \caption{Energy evolution during our simulations in the complete (blue solid line) and incomplete (black dotted line) scenarios using the finite fault inversion by Caltech Tectonics Observatory \cite{Ozgun2006}. Note that the scales are different on the left and right images. The time $t$ is given in seconds.}
  \label{fig:energiesCaltech}
\end{figure}

We performed also the same simulation but using the finite fault solution obtained by the Caltech team, \cite{Ozgun2006}, and the trigonometric scenario. The results concerning the kinetic and potential energies are presented in Figure \ref{fig:energiesCaltech}. It is interesting to note that the potential energy evolution in the Caltech scenario appears to be smoother especially after $t = 150$ s. However, the peaks corresponding to subfaults activation times are equally present in kinetic energies. The magnitudes of kinetic and potential energies predicted according to the Caltech inversion are approximatively 10 times higher than corresponding USGS results. 

\begin{figure}
  \centering
  \subfigure[USGS version]%
  {\includegraphics[width=0.495\textwidth]{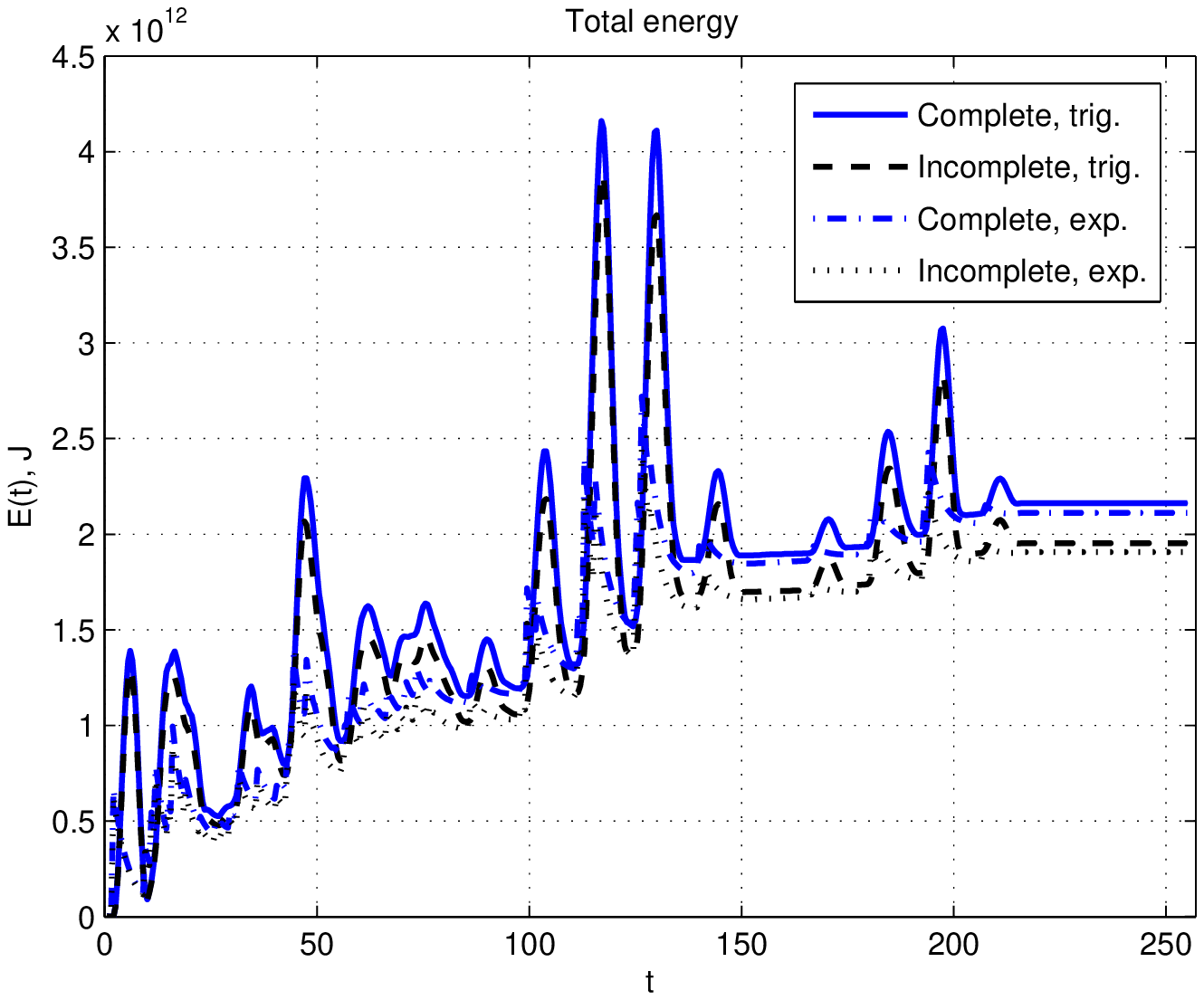}}
  \label{fig:totalEnergies}
  \subfigure[Caltech version]%
  {\includegraphics[width=0.495\textwidth]{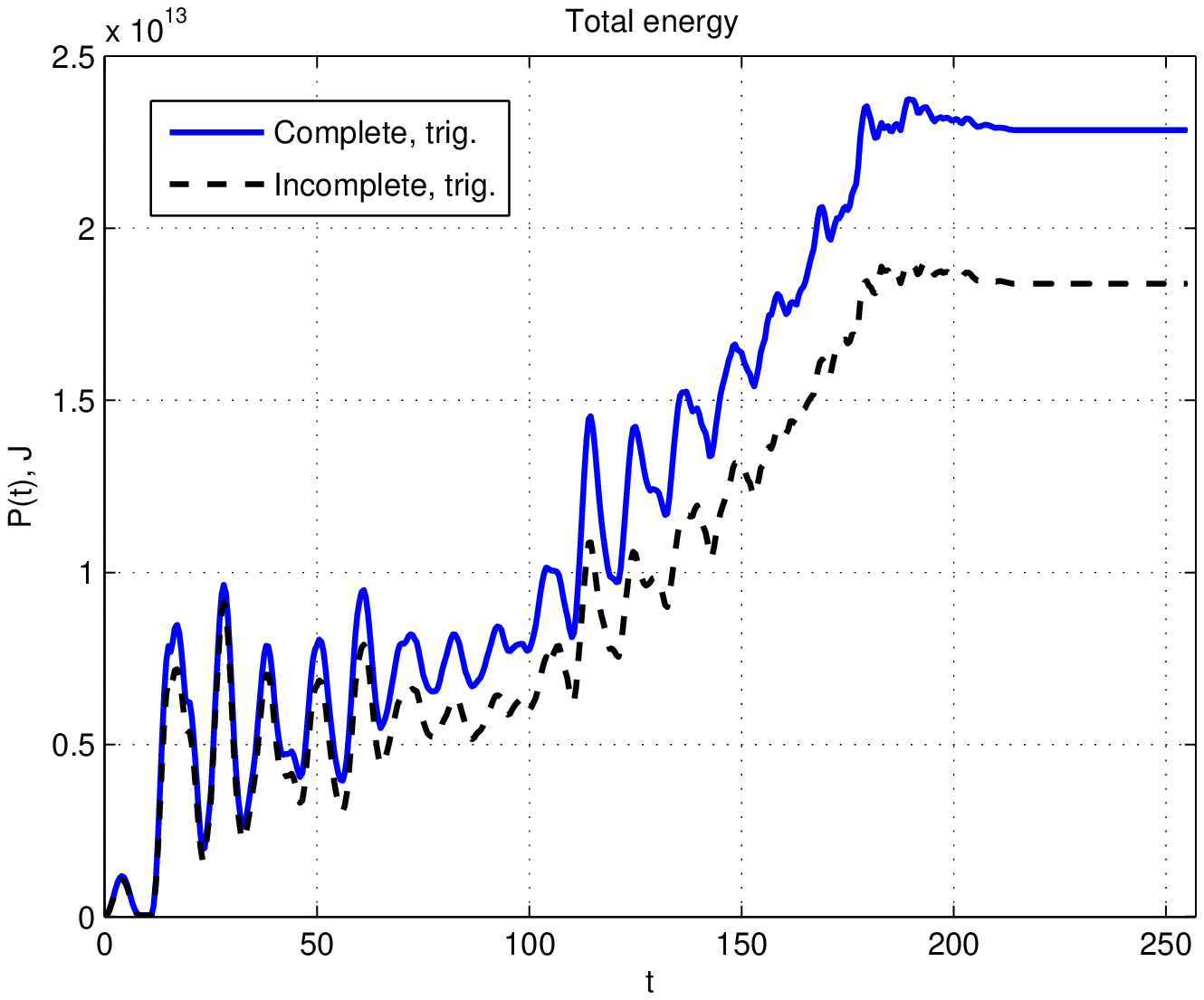}}
  \caption{The total energy evolution predicted according to two different versions of the finite fault inversion. Note the different vertical scales on left ($4.5\times 10^{12}$) and right ($2.5\times 10^{13}$) images. The time $t$ is given in seconds.}
  \label{fig:TotalE}
\end{figure}

We have to underline that inversions performed by the two different finite fault algorithms lead to very different results. Our method is operational with both of them. It is particularly interesting to compare the total energies evolution predicted by USGS and Caltech inversions. These results are presented in Figure \ref{fig:TotalE}. The Caltech version of the rupturing process generates a tsunami wave with much higher energy (computed after the end of the bottom motion when a tsunami enters in the propagation regime):
\begin{itemize}
  \item USGS, complete trigonometric: $E_c = 2.16\times 10^{12}$ J,
  \item USGS, complete exponential: $E_c = 2.11\times 10^{12}$ J,
  \item USGS, incomplete trigonometric: $E_i = 1.95\times 10^{12}$ J,
  \item USGS, incomplete exponential: $E_i = 1.90\times 10^{12}$ J,
  \item Caltech, complete trigonometric: $E_c = 2.28\times 10^{13}$ J,
  \item Caltech, incomplete trigonometric: $E_c = 1.83\times 10^{13}$ J.
\end{itemize}
This result is not surprising since according to the USGS solution the Java 2006 event magnitude is $M_w = 7.7$ and according to the Caltech team, $M_w = 7.9$, which means a huge difference since the scale is logarithmic.

\begin{remark}
We would like to note an interesting property. For linear waves it can be rigorously shown the exact equipartition property between the kinetic and potential energies \cite{Whitham1999}. When the nonlinearities are included, this equidistribution property is only approximate. One can observe that at the final time in our simulations the kinetic and potential energies are already of the same order of magnitude regardless the employed bottom motion (see Figures \ref{fig:energies} \& \ref{fig:energiesCaltech}). Moreover, both curves $K(t)$ and $\Pi(t)$ continue to tend to the equilibrium state according to the theoretical predictions of the water wave theory, \cite{Whitham1999}.
\end{remark}

Despite some local attenuation effects of horizontal displacements on the free surface elevation, the complete generation scenario produces a tsunami wave with more important energy content. More precisely, our computations show that the horizontal displacements contribute about $10 \%$ into the total tsunami energy balance in the USGS scenario (this value is consistent with our previous results concerning the differences in wave amplitudes in Figure \ref{fig:diffHV}). This result is even more flagrant for the Caltech version which ascribes $24\% $ of the energy to horizontal displacements. The free surface amplitudes in this case should differ as well by the same order of magnitude. As we already noted, the difference between trigonometric and exponential scenarios is negligible.

\begin{figure}
  \centering
  \includegraphics[width=0.9\textwidth]{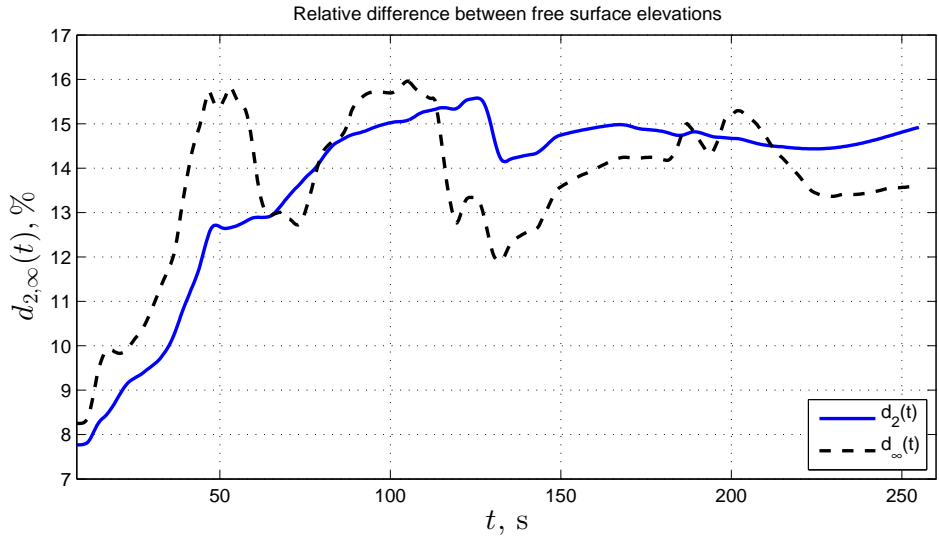}
  \caption{Relative difference between free surface elevations computed according to the complete and incomplete scenarios. For more details see equation \eqref{eq:d2inf}. The horizontal axis is given in seconds, while the vertical scale is a percentage. The blue solid line (\textcolor{blue}{---}) corresponds to the measure $d_2(t)$. The black dashed line ($---$) represents $d_\infty(t)$.}
  \label{fig:reldiff}
\end{figure}

Up to now the contribution of horizontal displacements has been quantified in terms of the bottom deformation (Figure \ref{fig:hordisp}), free surface elevation at a particular moment of time ($t = 220$ s, cf. Figure \ref{fig:diffHV}) and finally in terms of the wave energy (Figures \ref{fig:energies} -- \ref{fig:TotalE}). However, the wave energy cannot be directly measured in practice. Consequently, we continue to quantify the differences between complete and incomplete approaches in terms of the waveforms. Namely, we compute also the following relative measures of the waveforms difference which act on the whole computed free surface:
\begin{equation}\label{eq:d2inf}
  d_2(t) := \frac{||\eta_c(\x,t) - \eta_i(\x,t)||_{L_2(\R^2)}}{||\eta_i(\x,t)||_{L_2(\R^2)}}, \qquad
  d_\infty(t) := \frac{||\eta_c(\x,t) - \eta_i(\x,t)||_{L_\infty(\R^2)}}{||\eta_i(\x,t)||_{L_\infty(\R^2)}}.
\end{equation}
The simulation results are presented on Figure \ref{fig:reldiff}. One can see that both measures grow up to 15\% and then oscillate around this level, at least during the simulation time. This result is in agreement with previous measurements of the horizontal displacements contribution based on the wave energy and the bottom deformation which were of the order of 10\%. We note also that the curve $d_2(t)$ has a more regular behaviour than $d_\infty(t)$ since it is based on integral characteristics of the difference while the latter focusses on the characteristics of the extreme values.

It is also interesting to compare the tsunami energy with the energy of the underlying seismic event. The USGS Energy and Broadband solution indicates that the radiated seismic energy of the Java 2006 earthquake is equal to $3.2\times 10^{14}$ J. Hence, according to our computations with the USGS complete generation approach and the trigonometric scenario, about $0.68 \%$ of the seismic energy was transmitted to the tsunami wave (this portion raises to $7.14 \%$ for the Caltech version). The ratio of the total tsunami and seismic radiated energies can be used as a measure of the tsunami generation \emph{efficiency} of a specific earthquake. The seismic radiated energy may not be the most appropriate parameter to use, but it has an advantage to be relatively robust and easily observable in contrast to the more relevant seismic fracture energy, \cite{Venkataraman2004, Kanamori2006}.

This parameter can also be estimated for the Great Sumatra-Andaman earthquake of December 26, 2004. The total energy of the Great Indian Ocean tsunami 2004 was estimated by T.~Lay \emph{et al.} (2005), \cite{Lay}, to be equal to $4.2\times 10^{15}$ J. According to the same USGS Energy and Broadband solution the radiated seismic energy was equal to $1.1\times 10^{17}$ J. Thus, the tsunamigenic efficiency of the Great Sumatra-Andaman earthquake is $3.81$\% which is bigger than that of Java 2006 event but remains in the same order of magnitude around 1\%. It is possible that we do not still take into account all important factors in Java 2006 tsunami genesis. More examples of transmission of seismic energy to tsunami energy can be found e.g. in \cite{Lovholt2006}.

\section{Discussion and conclusions}\label{sec:concl}

In the present study the process of tsunami generation is further investigated. The current tsunamigenesis model relies on a combination of the finite fault solution,\cite{Ji2002}, and a recently proposed Weakly Nonlinear (WN) solver for the water wave problem with moving bottom, \cite{Dutykh2010a}. Consequently, in our model we incorporate recent advances in seismology and computational hydrodynamics.

This study is focused on the role of the horizontal displacements in the real world tsunami genesis process. By our intuition we know that a horizontal motion of the flat bottom will not cause any significant disturbance on the free surface. However, the real bathymetry is far from being flat. The question which arises naturally is how to quantify the effect of horizontal co-seismic displacements during real world events.

The primary goal of this study was to propose relatively simple, efficient and accurate procedures to model tsunami generation process in realistic environments. Special emphasis was payed to the role of horizontal displacements which should be also taken into account. Thus, the kinematics of horizontal co-seismic displacements were reconstructed and their effect on free surface motion was quantified. The evolution of kinetic and potential energies were also investigated in our study. In the case of July 17, 2006 Java event our simulations indicate that 10\% of the energy input can be seemingly ascribed to effects of the horizontal bottom motion. This portion increases considerably if we switch to the scenario proposed by Caltech Tectonics Observatory \cite{Ozgun2006}.

The results presented in this study do not still explain the reasons for the extreme run-up values caused by July 17, 2006 Java tsunami, \cite{Fritz2007}. At least we hope that the proposed methodology illustrated on this important real world event will be proved helpful in future studies. In our opinion a successful theory should incorporate also other generation mechanisms such as local landslides/slumps which are subject to large uncertainties at the current stage of our understanding. Until now there are no detailed images of the seabed which could support or disprove this assumption. In this study we succeeded to quantify the significance of horizontal displacements for the tsunami generation. 

\appendix

\section{Applications to some recent tsunami events}

In this Appendix we apply the techniques described in this manuscript to two recent significant events. We compute the energy transmission from the corresponding seismic event to the resulting tsunami wave with and without horizontal displacements.  Throughout this Appendix we use the trigonometric scenario and the finite fault solutions produced by USGS.

\begin{remark}
In later versions of the finite fault solution the activation and rise time, as well as the seismic moment are specified for each sub-fault. Consequently, below we use this information to produce more accurate bottom kinematics. It allows us to compute also the seismic moment rate function\footnote{The total released seismic moment is given by the integral of this function over the rupture duration (time).}.
\end{remark}

\subsection{Mentawai 2010 tsunami}

\begin{figure}
  \centering
  \subfigure[Total energy]%
  {\includegraphics[width=0.49\textwidth]{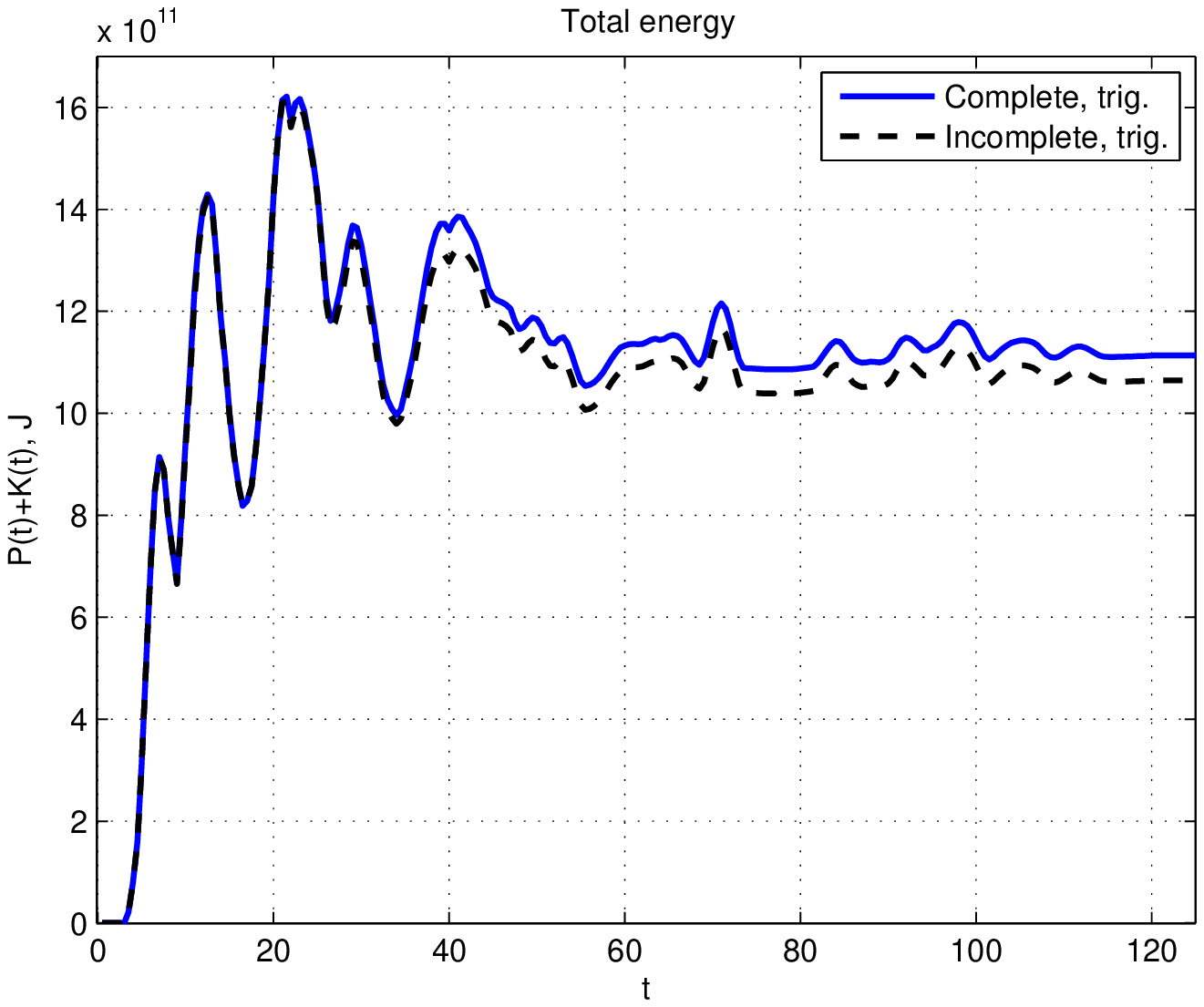}}
  \subfigure[Moment rate]%
  {\includegraphics[width=0.49\textwidth]{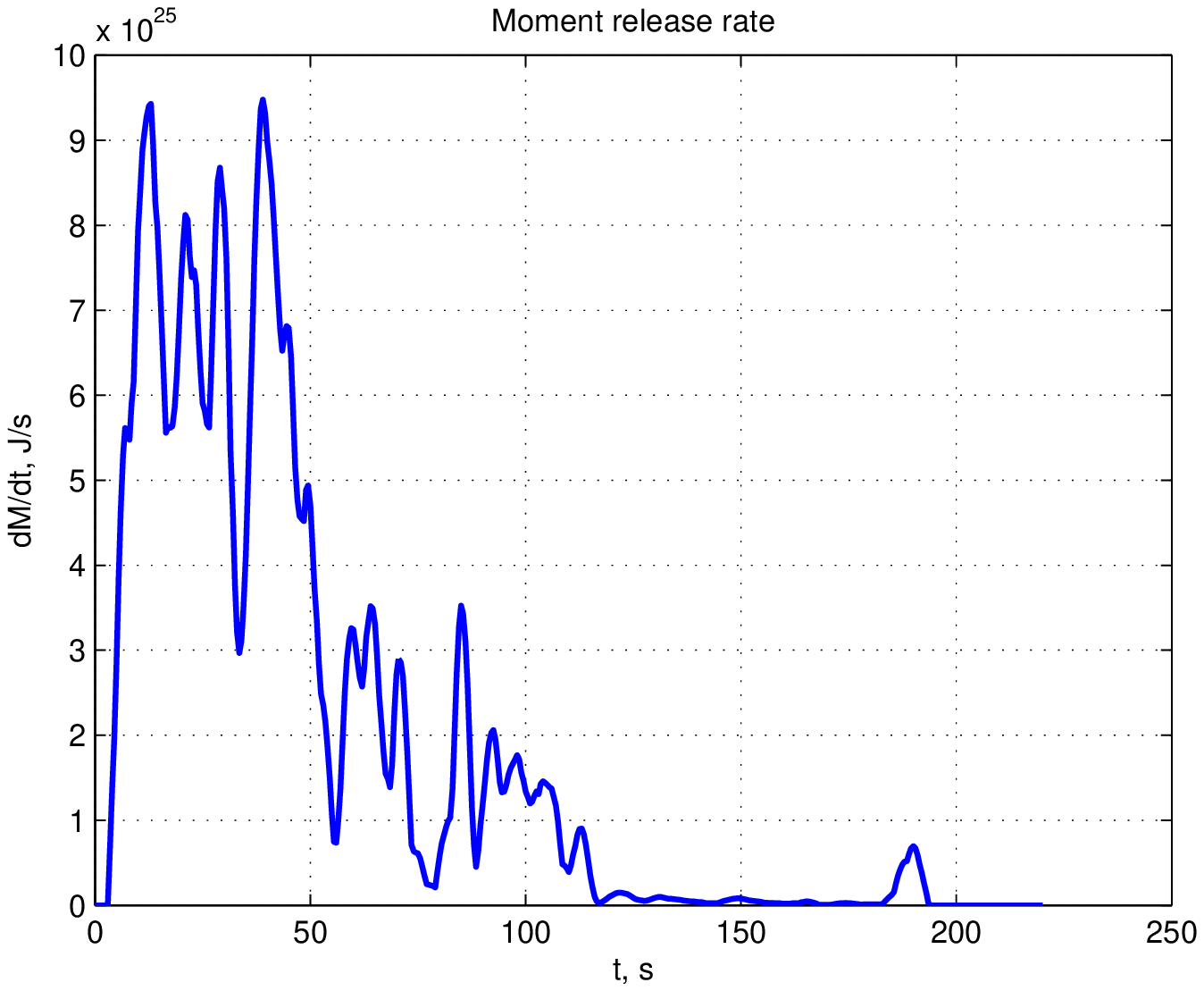}}
  \caption{Total tsunami energy and seismic moment rate function for the Mentawai 2010 event. The time $t$ is given in seconds.}
  \label{fig:Mentawai}
\end{figure}

In October 25, 2010 a small portion of the subduction zone seaward of the Mentawai islands was ruptured by an earthquake of magnitude $M_w = 7.7$, \cite{Lay2011}. This earthquake generated a tsunami wave with run-up values ranging from 3 to 9 meters (with even larger values at some places). On Pagai islands this tsunami caused more than 400 victims. This earthquake is characterized by $10^\circ$ dip angle, a slow rupture velocity ($\approx 1.5$ km/s) which propagated during about 100 s over 100 km long source region. For our simulation we used the finite fault solution produced by USGS \cite{Hayes2010}. On Figure \ref{fig:Mentawai} we show the evolution of the total tsunami energy to be compared with the seismic moment release rate function.

Our computations give the following estimations:
\begin{itemize}
  \item Complete scenario: $E_c = 1.11\times 10^{12}$ J,
  \item Incomplete scenario: $E_i = 1.06\times 10^{12}$ J.
\end{itemize}
Consequently, for this event only $4.5 \%$ of the total energy is due to horizontal displacements. According to the USGS energy and broadband solution the radiated seismic energy is $E_s = 1.4\times 10^{15}$ J. The tsunami energy constitutes $0.08 \%$ of the radiated seismic energy. This surprising result can be explained by the fact that a big portion of co-seismic displacements occured on the land which did not fortunately contribute to the tsunami generation process.

\subsection{March 11, 2011 Japan earthquake and tsunami}

This tsunami event was caused by an undersea megathrust earthquake of magnitude $M_w = 9.0$ off the coast of Japan. Officially this earthquake was named the \emph{Great East Japan Earthquake}. The Japanese National Police Agency has confirmed more than 13,000 deaths caused both by the earthquake and especially by the tsunami. Entire towns were devastated. The local infrastructure including the Fukushima Nuclear Power Plant were heavily affected with consequences which are widely known.

\begin{figure}
  \centering
  \subfigure[Total energy]%
  {\includegraphics[width=0.49\textwidth]{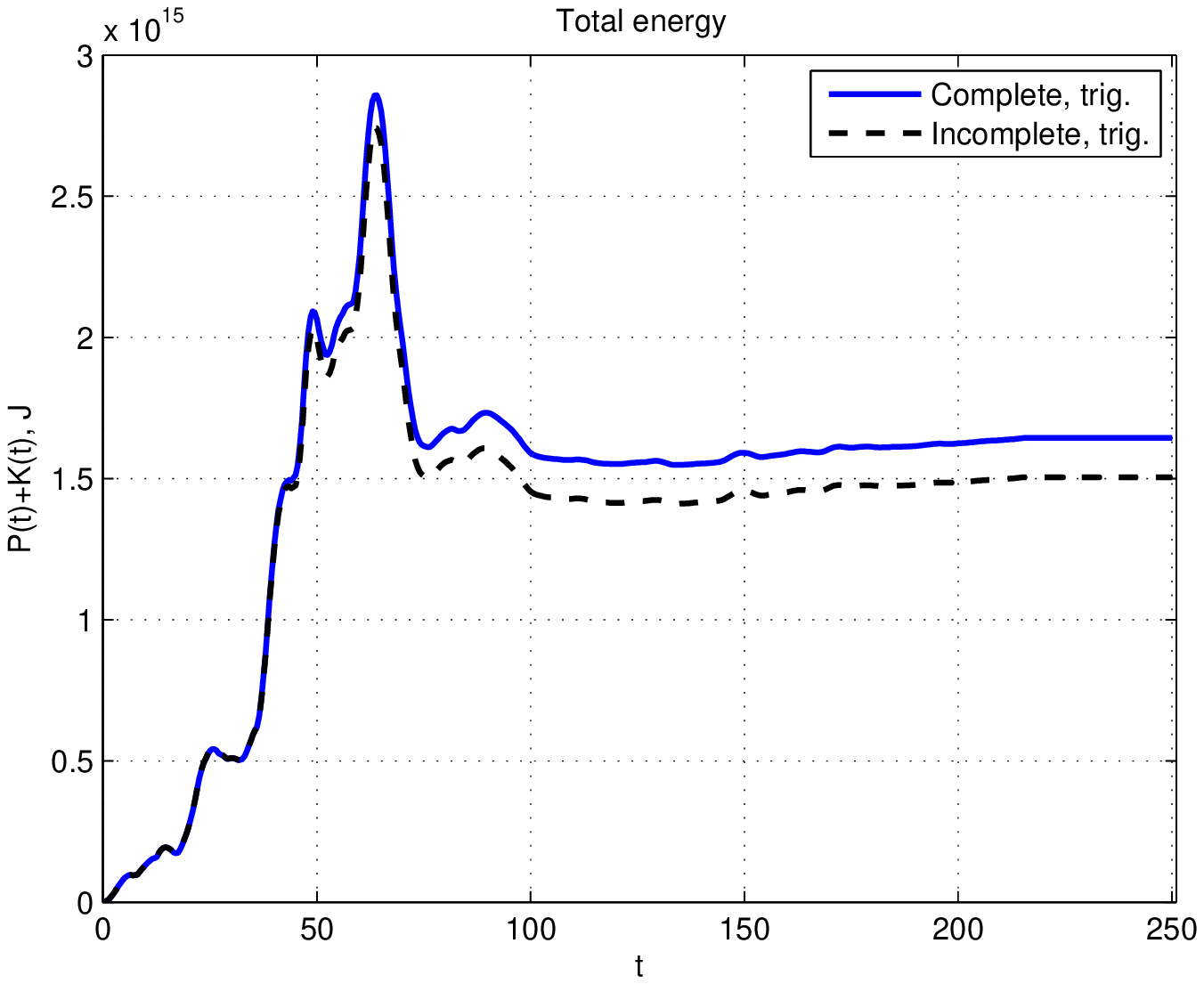}}
  \subfigure[Moment rate]%
  {\includegraphics[width=0.49\textwidth]{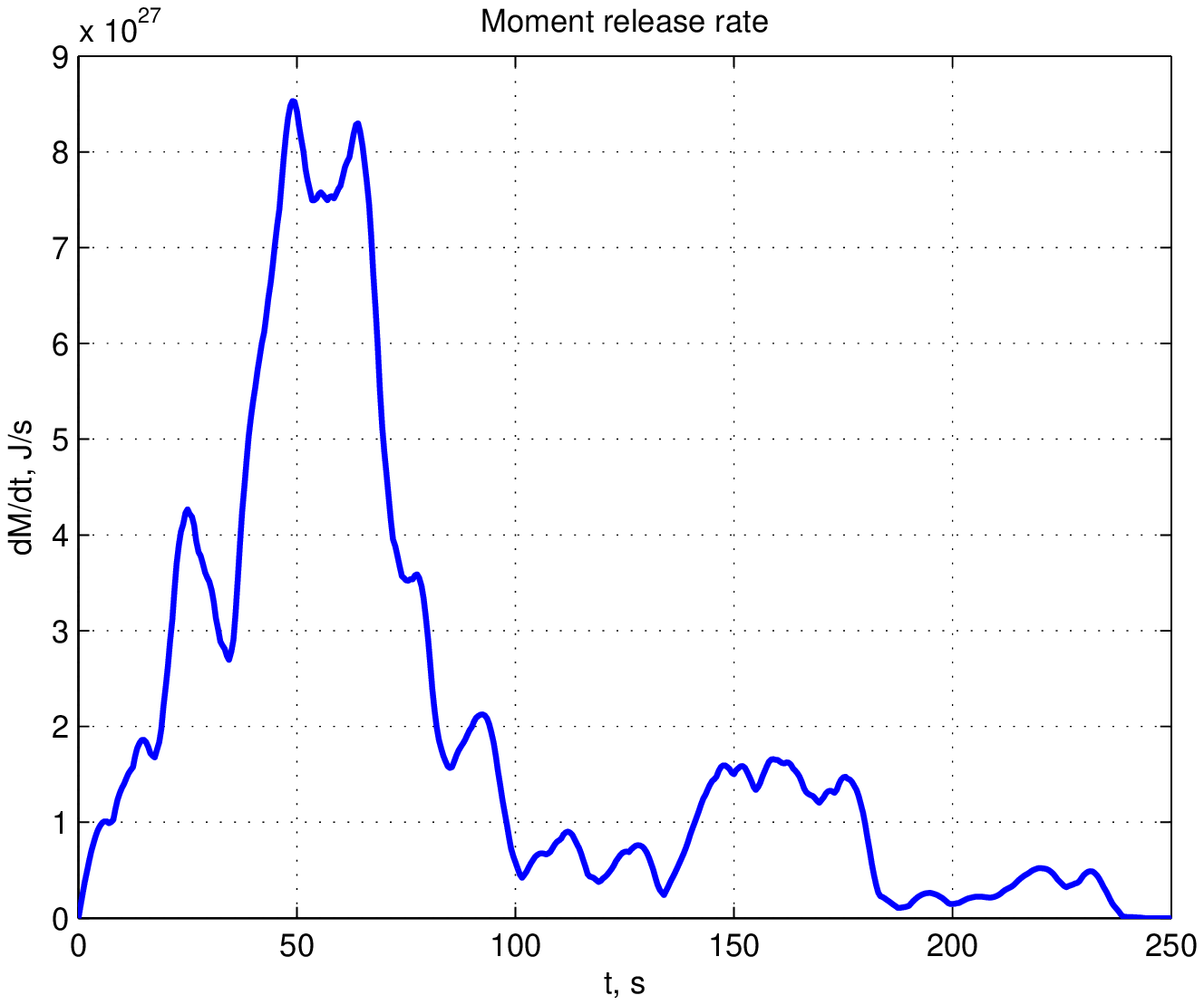}}
  \caption{Total tsunami energy and seismic moment rate function for the Great East Japan Earthquake. The time $t$ is given in seconds.}
  \label{fig:Japan}
\end{figure}

We use again the finite fault inversion performed at the USGS \cite{Hayes2011}. The tsunami total energy evolution along with the moment rate function are represented on Figure \ref{fig:Japan}. At the end of the simulation we obtain the following result:
\begin{itemize}
  \item Complete scenario: $E_c = 1.64\times 10^{15}$ J,
  \item Incomplete scenario: $E_i = 1.50\times 10^{15}$ J.
\end{itemize}
Henceforth, the contribution of horizontal displacements is estimated to be $9.2 \%$.
The radiated seismic energy is $E_s = 1.9\times 10^{17}$ J according to the USGS energy and broadband solution. The total tsunami wave energy represents only $0.87 \%$ of the radiated seismic energy. This value is much lower than the corresponding value for the Great Sumatra-Andaman earthquake of December 26, 2004.

\section*{Acknowledgement}
D.~Dutykh acknowledges the support from French Agence Nationale de la Recherche, project MathOcean (Grant ANR-08-BLAN-0301-01) and Ulysses Program of the French Ministry of Foreign Affairs under the project 23725ZA. French-Russian collaboration is supported by CNRS PICS project No. 5607. L.~Chubarov and Yu.~Shokin acknowledge the support from the Russian Foundation for Basic Research under RFBR Project No. 10-05-91052 and from the Basic Program of Fundamental Research SB RAS Project No. IV.31.2.1.

We would like to thank Professors Didier Clamond and Fr\'ed\'eric Dias for very helpful discussions on numerical simulation of water waves. Special thanks go to Professor Costas Synolakis whose work on tsunami waves has also been the source of our inspiration. Finally we thank two anonymous referees for the time they spent to review our manuscript and whose comments helped us a lot to improve its quality.

\bibliography{biblio}
\bibliographystyle{plain}

\end{document}